\documentclass[a4paper,11pt]{article}
\usepackage{aaskaiid}
\usepackage{booktabs}
\usepackage{titlesec}
\usepackage{orcidlink}
\usepackage{microtype} 
\usepackage[utf8]{inputenc} 

\titleformat{\paragraph}
{\normalfont\normalsize\bfseries}{\theparagraph}{1em}{}
\titlespacing*{\paragraph}
{0pt}{3.25ex plus 1ex minus .2ex}{1.5ex plus .2ex}

\title{Source Finding and Characterisation for SKAO Science}
\ShortTitle{Source Finding and Characterisation}

\author[1]{Sabyasachi Pal\orcidlink{0000-0003-2325-8509}}
\ShortName{Pal et al.} 
\author[1]{Souvik Manik\orcidlink{0000-0002-6794-7405}}
\author[2]{M Carmen Toribio\orcidlink{0000-0001-8063-2881}}

\author[3,7]{Geferson Lucatelli\orcidlink{0000-0002-2410-1776}}
\author[4,5]{Omkar Bait\orcidlink{0000-0003-2722-8841}}
\author[6]{Simone Riggi\orcidlink{0000-0001-6368-8330}}
\author[7]{Antxon Alberdi\orcidlink{0000-0002-9371-1033}}
\author[8]{Philippa Hartley}
\author[7]{Javier Moldon\orcidlink{0000-0002-8079-7608}}
\author[9]{Mamta Pandey-Pommier\orcidlink{0000-0001-5829-1099}}
\author[3]{Rob Beswick}

\affiliation[1]{Department of Pure and Applied Sciences, Midnapore City College, West Bengal, India}
\emailAdd{sabya.pal@gmail.com}
\affiliation[2]{Department of Space, Earth and Environment, Chalmers University of Technology, Onsala Space Observatory, SE-439 92 Onsala, Sweden}
\affiliation[3]{Jodrell Bank Centre for Astrophysics, School of Physics and Astronomy, The University of Manchester, Manchester M13 9PL, UK}
\affiliation[4]{National Radio Astronomy Observatory, 520 Edgemont Road, Charlottesville, VA 22903, USA}
\affiliation[5]{The NSF-Simons AI Institute for Cosmic Origins, USA, 201 E. 24th Street, POB 4.102, Austin, Texas 78712-1229} 
\affiliation[6]{INAF - Osservatorio Astrofisico di Catania, Via Santa Sofia 78, 95123 Catania, Italy}%

\affiliation[7]{Instituto de Astrof\'isica de Andaluc\'ia (IAA-CSIC), Glorieta de la Astronom\'ia s/n, 18008 Granada, Spain}
\affiliation[8]{SKA Organisation, Jodrell Bank Observatory, Cheshire, SK11 9FT, UK}
\affiliation[9]{Pole Scientific, University Catholic of Lyon, Campus Saint-Paul, 10 place des Archives 69288, Lyon Cedex 02, France}

\abstract{The advancements in highly sensitive and powerful radio telescopes, including the Square Kilometre Array Observatory (SKAO) and its precursors, MeerKAT, ASKAP, MWA, and HERA, will enable us to create the deepest radio images of the sky. However, due to the sheer scale of the datasets, manually classifying and analyzing this data is computationally expensive, time-consuming, and laborious. Therefore, the development of automated algorithms to detect and classify complex morphological radio sources from large astronomical surveys is the need of the time. In this chapter, we examine the recent advancements and challenges in source-finding techniques triggered by the analysis of SKAO precursor data and the SKA Data Challenges, both in spectral and continuum modes, along with the growing demand for computational resources and automated source detection methods using machine learning (ML) algorithms for the identification and characterisation of new populations of sources, addressing their complex and diffuse morphologies, as well as transient nature. Additionally, we discuss the critical factors affecting the quality and limitations of automated source-finding techniques, including artefacts from residual continuum emission, sidelobes, radio frequency interference (RFI), technical failures, calibration issues, flagging methods and false positives. With the availability of the full operational phase of the SKAO, continued advancements in source detection algorithms and computational infrastructure will be essential to fully exploit its scientific potential.}

\begin{document}
\maketitle

\section{Introduction}
\label{sec:intro_source_finding}

The Square Kilometre Array Observatory (SKAO) is poised to transform radio astronomy, delivering surveys with unprecedented depth, angular resolution, and sensitivity. These surveys will reveal billions of individual radio sources, from compact star-forming galaxies and active galactic nuclei (AGN) to vast diffuse structures such as relics and halos. Turning these data into scientific discovery depends critically on two linked processes: the automated finding of sources in survey images and the subsequent classification of their morphologies. Detection ensures that sources are reliably identified against complex noise backgrounds, while classification assigns physical meaning by distinguishing between structures, evolutionary stages, and emission mechanisms. Together, source finding and morphological classification form the backbone of SKAO data science.

Even this most fundamental task of \textit{source finding} has evolved into a major scientific and computational challenge in the SKAO era. The unprecedented sensitivity, angular resolution, and field of view of the SKAO and its pathfinder facilities, such as the Low-Frequency Array (LOFAR), the MeerKAT telescope, the Australian Square Kilometre Array Pathfinder (ASKAP), the upgraded Giant Metrewave Radio Telescope (uGMRT), the Very Large Array (VLA) and e-MERLIN, have ushered in an era of data abundance. These instruments collectively produce petabyte-scale data products, marking a shift from manually curated observations toward fully automated, data-intensive science.

Over the past decade, deep, wide-area continuum surveys such as the LOFAR Two-Metre Sky Survey (LoTSS; \citealt{Shimwell17, Shimwell22}), the MeerKAT International GHz Tiered Extragalactic Exploration (MIGHTEE; \citealt{MIGHTEE, Hale25}), and the Evolutionary Map of the Universe (EMU; \citealt{Norris21}) have increased the number of known radio sources by several orders of magnitude compared to earlier surveys like The NRAO VLA Sky Survey (NVSS; \citealt{Condon_1998}) and Faint Images of the Radio Sky at Twenty Centimeters (FIRST; \citealt{Becker95}), which each contained about $10^6$ sources. SKA Phase~I alone is expected to detect tens to hundreds of millions of sources. This growth introduces severe challenges for catalog creation, since every detected source must be characterized in terms of flux density, position, morphology, and spectral properties. Furthermore, the increasing diversity of morphologies, from compact AGN cores and star-forming galaxies to diffuse halos and relics, complicates automated detection even further.

The challenge is not merely one of scale but also of complexity. Radio interferometric imaging involves calibration, deconvolution, and mosaicking steps that can introduce artefacts, correlated noise, and position-dependent sensitivity. These effects can mimic or obscure real signals, making robust extraction difficult. As data volumes continue to grow, manual inspection and cleaning become infeasible. The field has therefore entered the \textit{big data paradigm} of astronomy, where automated, statistically rigorous, and computationally scalable methods are essential \citep{Whiting2012, riggi_2016}.

Reliable and complete source catalogues are fundamental to the scientific validity of downstream analyses. Incompleteness or systematic biases can propagate into derived quantities such as source counts, luminosity functions, and clustering statistics, ultimately affecting cosmological and galaxy evolution studies \citep{Hancock2012, hale_2019}. The ideal source finder must therefore optimize three key metrics: \textit{completeness} (the fraction of real sources detected), \textit{reliability} (the fraction of detections that are genuine), and \textit{parameter accuracy} (the fidelity of measured fluxes, positions, sizes, etc.). Balancing these simultaneously is a non-trivial problem, particularly for faint or diffuse sources near the noise level.

The evolution of automated source finding has mirrored advances in radio astronomy. Early tools such as \textsc{SAD} and \textsc{IMSAD} provided simple Gaussian fitting for compact sources but were limited by noise and extended emission. Later generations, \textsc{SExtractor} \citep{bertin1996sextractor}, \textsc{Aegean} \citep{Hancock2012}, \textsc{PyBDSF} \citep{Mohan2015}, and \textsc{Selavy} \citep{Whiting2012}, introduced adaptive thresholding, background estimation, and multi-component modeling. These algorithms, now integral to SKA precursor pipelines, incorporate multi-scale decomposition, island-based detection, and local noise estimation. However, they still assume Gaussian noise and simple morphologies, assumptions that often fail in the presence of direction-dependent calibration errors, residual sidelobes, or diffuse emission.

The challenge is even greater for spectral-line data cubes, where the third (frequency or velocity) dimension requires source finders capable of handling spatial-spectral correlations. Tools such as \textsc{SoFiA} (Source Finding Application; \citet{Serra2015}) and \textsc{DUCHAMP} \citep{whiting_2012} address these challenges using multidimensional thresholding, smoothing, and reliability analysis. These form the basis of pipelines for HI surveys like WALLABY using ASKAP \citep{westmeier2022wallaby} and LADUMA using MeerKAT \citep{blyth2016laduma}, and will be vital for future SKA-MID HI surveys. Nevertheless, they struggle with low signal-to-noise or irregular sources, motivating the integration of more adaptive, learning-based strategies.

A promising frontier is the incorporation of artificial intelligence (AI) and machine learning (ML) into source finding and classification. These approaches can capture complex, non-linear patterns and learn directly from data, bypassing many limitations of rule-based algorithms. Convolutional neural networks (CNNs), U-Net architectures, and generative models have shown success in source segmentation, denoising, and morphological classification \citep{bonaldi_2021, magro_2022, lao21}. Provided that sufficient representative training data exist, ML approaches can adapt to diverse morphologies, noise conditions, and instrumental effects. Active research continues on training data generation, transfer learning, and model interpretability, especially for the heterogeneous datasets expected from SKA Phase~I and II.

These advances must be supported by developments in data management and high-performance computing (HPC). The SKA Science Data Processor (SDP) will be among the largest HPC infrastructures ever built, processing exabytes of raw visibilities into science-ready catalogues. Efficient parallelization, distributed memory handling, and I/O optimization are integral to any SKA-scale source-finding framework. Equally crucial are reproducibility and transparency: pipelines must be automated yet auditable to ensure scientific trust in derived catalogues.

In summary, the SKAO era presents both an extraordinary opportunity and a profound computational challenge. The vast data volumes, complex noise environments, and morphological diversity of radio sources demand innovative, scalable, and intelligent algorithms. The field is evolving from classical detection and fitting toward hybrid methods that combine physical modelling, statistical inference, and AI-driven learning. The success of SKAO science will depend on such next-generation frameworks that can ensure catalogue integrity, reliability, and computational efficiency.

This chapter reviews the methods developed for source finding and classification, emphasizing lessons from current SKAO precursors and SKAO Data Challenges. It contrasts classical approaches with machine learning methods, assesses their strengths and limitations, and explores their integration into unified pipelines. The chapter concludes with a discussion on scalability, interpretability, and scientific reliability, and how these elements will shape SKA's transformational science.

\section{Source Finding Techniques for SKAO Science}

In the context of the SKAO and its pathfinder surveys, source finding has evolved from a relatively straightforward image-analysis task into one of the most computationally demanding challenges in modern astronomy. The large data volumes, high dynamic range, and complex source morphologies expected from SKAO observations demand algorithms that are not only statistically robust but also scalable and fully automated. Traditional source finders, developed primarily for smaller, shallower surveys, must now be re-evaluated and adapted to handle petabyte-scale datasets while maintaining high completeness and reliability. 

This section outlines the major developments in source-finding methodologies, beginning with classical detection and fitting algorithms, progressing to modern machine-learning approaches, and concluding with a comparative assessment of their performance and applicability to SKA-scale data processing.

\subsection{Classical Source Finding Techniques}
\label{subsec:classical_source_finding}
The rapid growth of modern radio surveys in both sky coverage and data volume has made automated, efficient, and statistically reliable source-finding algorithms indispensable. Contemporary source finders are expected to deliver high completeness and reliability while maintaining computational scalability across petabyte-scale datasets. Over the past two decades, a wide range of source-finding tools has been developed and optimized for different scientific and instrumental contexts. Among the most widely used are \textsc{Aegean} \citep{Hancock2012, Hancock2018}, the Astronomical Point source EXtractor (\textsc{APEX}; \citealt{makovoz_2005}), \textsc{blobcat} \citep{hales_2012}, the Curvature Threshold Extractor (\textsc{CuTEx}; \citealt{molinari_2011}), \textsc{Duchamp} \citep{Whiting2012}, and the IFCA Biparametric Adaptive Filter (\textsc{BAF}; \citealt{lopez_2012}) and Matched Filter (\textsc{MF}; \citealt{lopez_2006}) algorithms. Additional examples include \textsc{PyBDSF} \citep{Mohan2015}, the Python Source Extractor (\textsc{PySE}; \citealt{spreeuw_2010, swinbank_2015, carbone_2018}), \textsc{SAD} and its variant \textsc{HAPPY} \citep{Condon_1998, white_97}, \textsc{Selavy} \citep{whiting_2012}, \textsc{SExtractor} \citep{bertin_1996}, and the \textsc{SOURCE\_FIND} pipeline developed for AMI \citep{amic_2011}. More recent developments include \textsc{Caesar} \citep{riggi_2016, riggi_2019} and \textsc{ProFound} \citep{robotham_2018, hale_2019}. While many of these packages were originally developed for optical imaging (e.g. \textsc{APEX}, \textsc{CuTEx}, \textsc{ProFound}, and \textsc{SExtractor}), they have since been adapted for radio astronomical applications. In addition to two-dimensional (2D) continuum finders, specialized three-dimensional (3D) packages such as \textsc{SoFiA} \citep{Serra2015, koribalski_2020} are optimized for spectral-line data cubes and can also operate in 2D mode.

No single source finder performs optimally under all observing conditions or data regimes. Each tool is typically tuned for a particular type of source or image property \citep{hale_2019, bonaldi_2021}. Broadly, existing algorithms can be categorized into those optimized for compact sources and those tailored for extended or diffuse emission (see Table~\ref{tb:sf_characteristics}). Some also have specialized applications: for example, \textsc{blobcat} excels at handling linearly polarized data \citep{hales_2012}, \textsc{Duchamp} is widely used for H\,\textsc{i} emission detection \citep{Whiting2012}, \textsc{CuTEx} performs well in crowded or highly variable background conditions \citep{molinari_2011}, and \textsc{PySE} is particularly suited for transient searches \citep{fender_2007, haarlem_2013}. Several packages, including \textsc{Aegean}, \textsc{Selavy}, and \textsc{Caesar}, also offer multiprocessing or parallelized implementations to efficiently handle large-scale survey data. In parallel, novel frameworks based on machine learning \citep{bonaldi_2021, lao21, magro_2022} and citizen science efforts such as \textit{Radio Galaxy Zoo} \citep{banfield15, alger18} are being explored to complement these traditional approaches.

Most classical source-finding pipelines operate through three fundamental stages: (1) background and noise estimation, (2) island or region detection, and (3) component characterization or model fitting.

\begin{figure}[htb!]
\begin{center}
\includegraphics[width=\columnwidth,height=6cm]{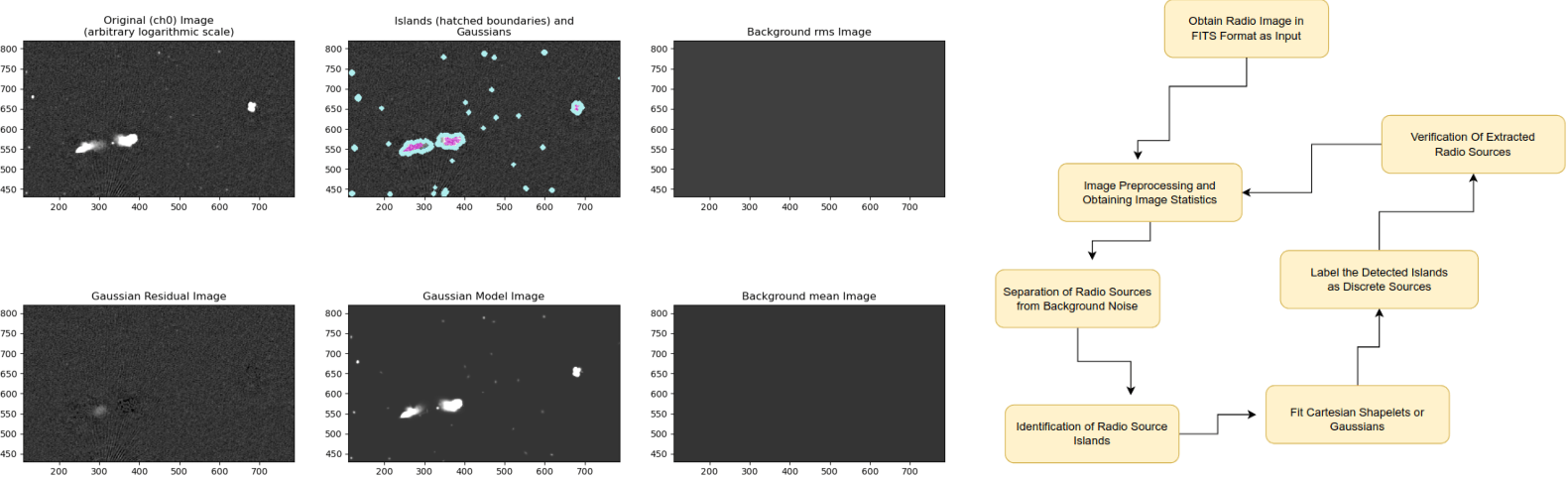}
\caption{Different stages of the process of characterisation of radio sources with widely used PyBDSF.} 
\end{center}
\end{figure}

\subsubsection{Background and noise estimation}
A reliable background model is crucial, as it directly affects the sensitivity and false detection rate of the pipeline. Many algorithms, such as \textsc{Aegean}, \textsc{PyBDSF}, and \textsc{Selavy}, use a sliding-box approach in which the local background and noise levels are estimated using neighboring pixels within a defined box size. The choice of box size is critical: too small a region leads to noise overestimation near bright sources, whereas too large a region can smooth over real spatial variations in the background \citep{huynh_2012}. Different statistical estimators are used across finders: for instance, \textsc{Aegean} uses the median background with the inter-quartile range (IQR) as a noise measure \citep{Hancock2012, Hancock2018}; \textsc{PyBDSF} relies on mean ($\mu$) and root-mean-square (RMS, $\sigma$) noise values \citep{Mohan2015}; while \textsc{Selavy} supports both $\mu$/$\sigma$ and median/MADFM (mean absolute deviation from the median) estimators \citep{whiting_2012}. Optical-origin codes such as \textsc{SExtractor} and \textsc{ProFound} apply $\kappa\sigma$-clipping and mode estimation across the image \citep{bertin_1996, robotham_2018}. The choice of estimator significantly influences detection completeness and reliability \citep{huynh_2012}.

In practice, the performance of these methods is highly sensitive to the parameter 
choices, including box size, clipping thresholds, and estimator selection, which often requires domain expertise to optimise for a given dataset. Default parameter 
settings may not generalise well across surveys with differing resolutions, 
sensitivities, and noise characteristics, potentially leading to systematic biases 
in source detection and characterisation. This dependence on expert tuning presents 
a challenge for fully automated pipelines, highlighting the need for robust, 
adaptive methods that can minimise manual intervention while maintaining consistent 
performance across heterogeneous datasets.

\subsubsection{Island detection}
Once the background is estimated, candidate emission regions, often termed ``islands'', are identified. Threshold-based detection remains the most common approach, in which contiguous pixels above a specified multiple of the noise level are grouped into islands. Variants of this approach are employed in \textsc{Duchamp}, \textsc{ProFound}, \textsc{Selavy}, and \textsc{SExtractor} \citep{whiting_2012, robotham_2018, bertin_1996}. Some packages employ more sophisticated region-growing techniques such as flood-fill algorithms (\textsc{Aegean}, \textsc{PyBDSF}, \textsc{blobcat}, and \textsc{Caesar}) or iterative dilation methods inspired by Kron/Petrosian apertures (\textsc{ProFound}; \citealt{petrosian_1976,kron_1980}). In certain cases, watershed deblending is used to separate overlapping sources by following the topographic ``valleys'' in brightness distribution.

\subsubsection{Component modeling}
After island detection, the emission is decomposed into individual source components. The simplest approach identifies brightness peaks within each island and fits them with 2D elliptical Gaussians, an appropriate model for compact sources that resemble the synthesized beam shape. This strategy is used by \textsc{Duchamp} and \textsc{Selavy} \citep{whiting_2012}. More complex finders such as \textsc{PyBDSF} and \textsc{PySE} fit multiple Gaussians per island to better represent extended or multi-component sources \citep{Mohan2015, swinbank_2015}. Algorithms like \textsc{Aegean} and \textsc{Caesar} incorporate curvature maps or local peak searches to constrain Gaussian fits more robustly \citep{Hancock2012, Hancock2018, riggi_2016}. \textsc{Caesar} \citep{Riggi19}, in particular, combines peak identification with watershed-based segmentation and applies additional filtering (e.g. wavelet, hierarchical clustering, or active contour methods) to extract faint, diffuse, or irregular structures, making it one of the most flexible frameworks for extended emission.

Overall, while classical source finders remain the workhorses of radio survey pipelines, their assumptions about noise, morphology, and source structure limit their applicability in the high-complexity data environment of the SKAO era. This motivates the transition toward hybrid and machine learning based approaches capable of operating at the required sensitivity, speed, and morphological diversity of next-generation surveys.

\subsection{3D Source Finding and Spectral Line Surveys}
For spectral-line observations, such as HI surveys, the problem of source finding extends into three dimensions: two spatial axes and one spectral (velocity or frequency) axis. In this case, the goal is to detect emission features across multiple frequency channels while accounting for correlated noise and line broadening due to galaxy rotation or turbulence. Classical 2D thresholding is inadequate for this task because real emission often appears as faint, contiguous structures in both spatial and spectral dimensions.

\subsubsection{HI emission source finding}
A key difference between source-finding in the continuum and in HI emission is that HI emission is relatively faint compared to the noise floor of the given image plane or spectral channel. Fortunately, HI astronomers are able to recover the HI emission through the increase in SNR from integrating across the full frequency range that the HI emission spans.  This means that the contiguity of emission across the frequency span provides useful information about the presence of an HI source and is typically crucial in detecting HI sources (and separating true source emission from radio frequency interference (RFI) false positives, for example).  

Specialized 3D source finders, such as \textsc{DUCHAMP} \citep{Whiting2012} and the \textsc{Source Finding Application} (\textsc{SoFiA}) \citep{Serra2015}, were developed to address this challenge. \textsc{SoFiA} is currently the main tool HI astronomers turn to for automated source finding and characterisation. As evident from SDC2, many of the teams have relied on SoFiA either as the singular tool used to perform source-finding and characterisation, or in combination with newer machine learning methods. \textsc{DUCHAMP} implements a multi-threshold approach with optional smoothing kernels to enhance signal detectability, while \textsc{SoFiA} offers a modular framework combining pre-filtering, detection, reliability analysis, and parametrization. The more recent \textsc{SoFiA2} \citep{sofia2} provides better scalability and performance for modern surveys such as WALLABY (ASKAP) and LADUMA (MeerKAT). These tools are critical for the upcoming SKA-MID HI surveys that aim to detect millions of galaxies in emission up to redshift $z \sim 1$. Despite their success, 3D finders also face difficulties in distinguishing faint extended structures from correlated noise, motivating the exploration of machine-learning-based approaches.

The strength of ML-based source finders lies in their ability to model non-Gaussian noise systematics that are inherent in real HI survey datasets. In recent times, several groups have achieved greater performance in source-finding through integrating machine learning based methods with more traditional tools such as SoFiA (e.g. \citealt{Barkai2023, hartley23, haakansson2023utilization}).  Recently, \citet{wang25} has also demonstrated the potential to use a machine learning-based method to streamline false positive rejection within the ASKAP WALLABY survey's source finding workflow with excellent efficiency.
 
\subsubsection{HI absorption source finding}
The challenge of automated HI absorption source finding is distinct from that of HI emission.  This is largely due to the typically narrower linewidths of the narrow component and the relatively weak but broad linewidth of the broadened component.  As such, the stability of the bandpass is crucial for accurately identifying an HI absorption line.

The MeerKAT MALS survey employs an automated method based on maximum likelihood optimization and iterative fitting of multiple Gaussian components within an absorption profile, with the final solution determined using Bayesian information criteria \citep{ngupta25}.

Similar to SoFiA—which is used to identify HI sources in emission—the ASKAP FLASH survey, a large untargeted HI absorption survey, employs FLASHfinder \citep{Allison12}, a multi-nested Bayesian Monte Carlo tool that simultaneously detects and fits HI absorption features. The performance of FLASHfinder as a production-ready, automated HI absorption line finder can be found in \citet{yoon25}.  Similar to the ASKAP WALLABY source finding workflow, visual inspection is still required to separate the false positives from the true detections.  As such, the team is working to develop more complex machine learning based solutions to address the non-Gaussian noise statistics that are present in the data products.

\vspace{0.4cm}

\begin{table}[htb]
   \caption{Summary of source finders design characteristics}
   \centering
   {\small
   \begin{tabular}{|l|c|c|c|c|}
      \hline
      \textbf{Source Finders} & \multicolumn{3}{c|}{\textbf{Source Type}} & \textbf{Multiprocessing} \\
      \cline{2-4}
                              & Compact & Extended & Diffuse & \\
      \hline
      PyBDSF                  & \checkmark & \checkmark &       &            \\ \hline
      Aegean                  & \checkmark &          &         & \checkmark \\ \hline
      APEX               & \checkmark &          &         &            \\ \hline
      ProFound           & \checkmark & \checkmark & \checkmark &          \\ \hline
      blobcat        & \checkmark & \checkmark &       &  \checkmark  \\ \hline
      Caesar                  & \checkmark & \checkmark & \checkmark & \checkmark \\ \hline
      CuTEx              & \checkmark &          &         &            \\ \hline
      IFCA BAF/MF             &            & \checkmark &       &            \\ \hline
      Selavy                  & \checkmark &          &         & \checkmark \\ \hline
      PySE                    & \checkmark &          &         &            \\ \hline
      SAD                     & \checkmark &          &         &            \\ \hline
      SExtractor        &            & \checkmark &       &            \\ \hline
      SOURCE\_FIND            &            & \checkmark &       &            \\ \hline
   \end{tabular}
   }
   \vspace{0.5em}
   \label{tb:sf_characteristics}
\end{table}

\subsection{Machine Learning in Source Finding}
With the advent of large data volumes and increasingly complex source morphologies, machine learning has emerged as a powerful alternative for automated source detection and classification. ML approaches, particularly deep learning (DL), enable the identification of non-linear patterns and subtle features that are often undetectable by traditional algorithms. In radio astronomy, CNNs and U-Net architectures have been successfully applied to tasks such as source segmentation, denoising, and morphological classification \citep{Akeret_2017,Aniyan17,Lukic2019,wu2019claran}. These models learn directly from labeled datasets, automatically distinguishing real sources from noise or imaging artefacts.

ML-based source finding methods can be broadly divided into supervised and unsupervised approaches, each with distinct advantages and limitations. Supervised methods require labeled training data in which source positions, extents, and morphologies are known, typically derived from simulations or cross-matched catalogs rather than direct ground truth in real observations. This introduces several challenges: the ground truth for real radio data is rarely known with certainty; simulations may not fully capture the complexity of real images, including correlated noise structures, calibration artefacts, and the morphological diversity of extended sources; and biases present in the training set risk propagating into derived catalogs \citep{Lukic2019, Riggi2023}. By contrast, unsupervised and self-supervised approaches learn representations directly from the data distribution without labeled examples, making them potentially more generalisable across heterogeneous surveys, though typically at the cost of less precise source localisation and harder quantitative validation.

The study by \citet{hopkins15} demonstrated that while traditional source finders perform well for compact sources, they lack sufficient robustness in extracting extended or diffuse emission. Deep learning architectures such as U-Net offer pixel-level segmentation capabilities, making them effective for detecting extended sources and diffuse structures that classical methods often miss. Similarly, hybrid methods that combine classical pre-processing (e.g., noise estimation or thresholding) with deep learning refinement have shown promising results in improving both completeness and reliability, particularly in crowded or high-noise environments \citep{Vafaei2019,Riggi2023}. Recent SKA Data Challenges and precursor surveys have provided valuable benchmarks for evaluating ML-based methods, allowing quantitative comparisons against classical approaches \citep{bonaldi_2021,hartley23}.

AI-based source finders such as \textsc{COSMODEEP} \citep{Gheller2018}, \textsc{DEEPSOURCE} \citep{Vafaei2019}, and \textsc{ConvoSource} \citep{Lukic2019}, employing custom CNN architectures, as well as more recent methods including \textsc{Astro R-CNN} \citep{Burke2019}, \textsc{Tiramisu} \citep{pino2021semantic}, \textsc{CAESAR-MRCNN/YOLO} \citep{Riggi2023}, \textsc{HeTu} \citep{Lao2023}, \textsc{RadioGalaxyNET} \citep{Gupta2024}, \textsc{YOLO-CIANNA} \citep{Cornu2024}, and \textsc{ContinUNet} \citep{ContinUNet} $-$ adopting either semantic segmentation architectures such as \textsc{U-Net} \citep{ronneberger2015u} or object-detection frameworks like \textsc{Mask R-CNN} \citep{he2017mask} and \textsc{YOLO} \citep{YOLO} $-$ have demonstrated high detection accuracies (often exceeding 90\%), particularly in recovering faint and diffuse sources that remain challenging for traditional algorithms.

Notably, \citet{ContinUNet} developed \textsc{ContinUNet}, a U-Net-based framework specifically designed for radio continuum source finding, demonstrating strong performance in recovering faint and extended emission across survey-like conditions.

\citet{Taran23} presented a novel deep learning framework that bypasses the traditional imaging step entirely by taking as input a low-dimensional vector of sampled visibility (uv-plane) data and directly outputting sky-source positions. They demonstrate with simulations (based on Atacama Large Millimeter/submillimeter Array; ALMA) that their ML-model achieves significantly higher completeness than standard image-plane source-finders (e.g., PyBDSF) in the low-signal-to-noise regime (S/N $\approx 1 - 10$) and is much faster ($\sim 30$x) since it avoids the full imaging pipeline. \citet{Drozdova24} take a different approach: they use simulated ALMA ``dirty'' radio-interferometric images and apply a conditional denoising diffusion probabilistic model (DDPM) to reconstruct sky models (i.e., clean images) and then perform source localization and flux estimation from those reconstructions. They show that their method delivers high completeness (over 90\%) even at S/N $\approx$ 2, and yields more accurate flux estimates than the traditional CLEAN + PyBDSF approach. Importantly, the stochastic nature of the DDPM also allows uncertainty quantification (in source location and flux density) for detected sources. 

Together, these works illustrate two complementary AI/ML-driven approaches for source finding for next-generation radio-interferometric data: one directly from visibilities, the other via improved image reconstruction prior to detection, both promising for the SKAO era where data volumes, faint-sources and imaging pipeline complexity pose major challenges for traditional approaches. Despite their advantages, ML-based methods depend heavily on the quality and diversity of their training datasets. Generating realistic, labeled simulations that capture the full range of SKAO data characteristics---including calibration residuals, instrumental noise, and sky complexity is a non-trivial task. Moreover, the interpretability and generalizability of deep learning models remain open challenges, underscoring the need for transparent and physically motivated architectures. 

\subsection{Comparative Assessment}
Traditional source-finding algorithms rely on deterministic rules such as noise thresholding, Gaussian fitting, and curvature-based detection. While these methods perform well for compact and isolated sources, they often struggle with complex, diffuse, or blended structures, especially in high-noise or crowded regions. Their performance is highly sensitive to parameter tuning, background estimation, and beam characteristics, making them less adaptable to the heterogeneity of modern radio surveys.

Quantitative comparisons of traditional source finders have been carried out in a number of studies. For instance, \citet{hopkins15} and \citet{Hancock2018} evaluated multiple algorithms (including \textsc{Aegean}, \textsc{PyBDSF}, and \textsc{Selavy}) using simulated and real radio data, demonstrating that completeness and reliability can vary by $\sim$10--20\% depending on source morphology, signal-to-noise ratio, and parameter choices. These studies demonstrate that no single algorithm performs optimally across all source types and survey configurations, particularly for extended and low-surface-brightness emission.

By contrast, ML approaches, particularly those based on deep learning, can learn complex non-linear relationships directly from the data, enabling more robust identification of faint, irregular, or overlapping sources. Recent works (e.g., \citealt{Lukic2019}; \citealt{wu2019claran}) show that ML-based methods can achieve improved completeness at low signal-to-noise levels, often outperforming traditional techniques by $\sim$10\% or more in challenging regimes, while maintaining comparable reliability. However, these gains depend on the availability of representative training data and careful validation to avoid biases.

As SKAO and its precursors push toward petabyte-scale imaging, integrating ML within automated pipelines has become increasingly essential for scalable, accurate, and efficient source detection and classification.

\section{Morphological Classification of Radio Sources}
\subsection{The Scientific Motivation}
Detecting sources is only the first step in the radio data analysis pipeline; the subsequent morphological classification is crucial for interpreting the underlying astrophysics. Radio morphology encodes key information about the physical mechanisms powering emission, such as relativistic jets from AGN, star formation (SF) processes, or large-scale shocks in galaxy clusters. Distinguishing between compact star-forming galaxies, Fanaroff–Riley type I/II radio galaxies, bent-tailed sources, and hybrid morphology radio sources (HyMoRS) offers insight into feedback processes, jet–environment interactions, and the co-evolution of galaxies and their central black holes. 

SKAO and its precursors (ASKAP, MeerKAT, LOFAR, uGMRT), morphological classification is not merely a cataloguing exercise; it underpins several of SKAO's primary science goals, from tracing cosmic magnetism and black hole growth to probing large-scale structure formation. Automated and reliable classification of the billions of sources expected in SKAO continuum surveys is therefore an essential component of future radio astronomy.

\subsection{Classical Approaches to Radio Source Classification}
Historically, the classification of radio sources has relied heavily on visual inspection and manual interpretation, supported by simple parametric descriptors such as flux ratios, spectral indices, symmetry measures, and Gaussian component fitting \citep{fanaroff_riley_1974, Condon_1998}. Early radio surveys such as NVSS and FIRST enabled the construction of large morphological catalogues, but these methods were inherently limited by human subjectivity, scalability, and sensitivity to faint extended emission.

Radio sources exhibit a remarkable diversity of morphologies, reflecting differences in their central engines, environments, and evolutionary stages. The most fundamental classification scheme, introduced by \citet{fanaroff_riley_1974}, divides extended radio galaxies into two broad categories: \textit{Fanaroff–Riley type~I} (FR~I) and \textit{type~II} (FR~II). FR~I sources display edge-darkened morphologies with brightness peaking near the core, typically associated with lower radio power and turbulent, decelerating jets. In contrast, FR~II sources are edge-brightened, with luminous terminal hotspots at the ends of collimated jets, corresponding to higher jet power and often more efficient particle acceleration. This dichotomy remains one of the cornerstones of radio galaxy classification and continues to inform models of jet dynamics and AGN feedback.

Beyond the canonical FR classes, a rich population of atypical and complex morphologies has been identified. These include HyMoRS, which simultaneously exhibits FR~I-like structure on one side of the nucleus and FR~II morphology on the other, suggesting asymmetric environments or jet instabilities \citep{gopal_krishna_2000, Gaw06, kapinska17, harwood20, kumari22, Manik2026}. Another prominent subclass comprises \textit{giant radio galaxies} (GRGs), whose projected sizes exceed 0.7~Mpc, making them among the largest single extragalactic structures known in the Universe \citep[e.g.][]{ishwara99, Dabhade2020, Delhaize2021, oei24, manik2024grs, manik2025grq, Charlton2025}. GRGs serve as probes of intergalactic magnetic fields, low-density environments, and long-term AGN activity cycles.

The \textit{Radio Galaxy Zoo} (RGZ) project \citep{banfield15}, a landmark citizen-science initiative, revolutionized morphological classification by engaging tens of thousands of volunteers to visually inspect radio sources from the FIRST \citep{Becker95} and Australia Telescope Large Area Survey (ATLAS) \citep{norris06, mao12}, cross-matched with Wide-field Infrared Survey Explorer (WISE) \citep{wright10} data. The RGZ effort produced a rich, labeled dataset of complex morphologies and provided invaluable training material for machine-learning models such as \textsc{CLARAN} \citep{wu2019claran}. However, manual or rule-based classification cannot meet the demands of SKAO-scale surveys, which will detect billions of sources spanning orders of magnitude in flux density and angular scale. Traditional approaches also struggle with composite or overlapping systems where multiple jets, lobes, or cores complicate identification.

Apart from the typical double-lobed radio galaxies that define the FR classification, numerous peculiar morphologies have been documented. \textit{Winged radio galaxies} (WRGs), which include \textit{X-shaped} (XRGs) and \textit{Z-shaped} (ZRGs) systems, display secondary lobes or ``wings'' that may arise from jet reorientation, backflow deflection, or black hole mergers \citep[e.g.][]{yang19, Bera20, bhukta22}. \textit{Double–double radio galaxies} (DDRGs) exhibit two distinct pairs of lobes aligned along the same axis, providing evidence for recurrent AGN activity \citep[e.g.][]{Konar2013}. 

Another notable category is that of \textit{bent-tailed (BT)} radio galaxies, which are subdivided into \textit{wide-angle tail (WAT)} and \textit{narrow-angle tail (NAT)} sources. These are typically located in dense cluster environments, where ram pressure from the intracluster medium bends the relativistic jets \citep[e.g.][]{Missaglia2019, Bh22tailed,Pal2023}. BT galaxies are valuable tracers of cluster dynamics and galaxy motions within the intracluster medium.

Recent surveys have also uncovered entirely new and rare morphologies. One striking example is the class of \textit{odd radio circles} (ORCs), enigmatic ring-like radio structures discovered in deep ASKAP observations \citep[e.g.][]{norris2021, Kumari2023,Lochner2023,Kumari2024_horseshoe, manik2025by}. Also see the chapter on the study of circular symmetric diffused radio sources with SKAO in \citet{SabyasachiPal02.2026.SKA}. Their origin remains under debate, with hypotheses ranging from AGN-driven shock fronts to spherical remnants of energetic outflows. Similarly, increasingly sensitive imaging has revealed asymmetric, fragmented, and hybrid structures that challenge existing classification schemes \citep{harwood20,kapinska17,Mingo_2019,Rudnick_2021, Kumari2024, Kumari2026}.

These discoveries collectively highlight the morphological richness of the radio sky and the limitations of traditional classification approaches. Manual inspection remains invaluable for discovering rare or peculiar sources, but it is insufficient for the data volumes anticipated in the SKAO era. The next generation of classification frameworks must, therefore, be automated, data-driven, and scalable, capable of recognizing both canonical and unconventional morphologies while maintaining scientific reliability across billions of detections.

Historically, morphology was studied by visual inspection, supported by simple quantitative measures such as flux ratios, axial symmetry, and fitted profiles. Campaigns like \textit{Radio Galaxy Zoo} demonstrated the power of citizen science in scaling up morphological classification, producing labeled datasets that remain invaluable for training and validation. You may find a chapter on the prospect of citizen science research with SKAO in \citet{Hota01.2026.SKA}. However, classical classification cannot meet the demands of SKAO data volumes. Visual inspection does not scale to billions of sources, and rule-based algorithms lack the flexibility to capture the diversity of real morphologies.

\subsubsection{Non-Parametric Source Characterisation}
\label{subsubsec:nonparametric_characterisation}
Before discussing parametric decomposition approaches, it is important to acknowledge the complementary role of non-parametric morphological metrics. These metrics quantify structural properties without assuming a specific mathematical form for the brightness distribution, making them valuable for characterising irregular, disturbed, or complex morphologies that may not be well-described by smooth analytical functions. In optical astronomy, the CAS system, measuring \textit{Concentration} (C), \textit{Asymmetry} (A), and \textit{Smoothness} or \textit{Clumpiness} (S), has proven effective for distinguishing between galaxy types and systems undergoing merging activity \citep{Conselice_2000, Lotz_2004}. This framework has been extended to include the \textit{Gini coefficient} (G), which quantifies the distribution of flux amongst pixels, and the second-order moment of the brightest 20 per cent of the flux ($M_{20}$), which traces the spatial distribution of the most luminous regions \citep{Lotz_2004, Lotz_2008}. Collectively, the CASGM system provides a robust, quantitative description of galaxy structure that is sensitive to mergers, interactions, and morphological disturbances.

Although these metrics have been predominantly developed in optical astronomy, their extension to radio wavelengths is straightforward and offers significant potential for characterising the diverse structures observed in radio sources. Concentration indices distinguish centrally dominated AGN cores from diffuse jets and extended star formation, asymmetry measurements quantify disturbances from mergers or AGN feedback, and smoothness parameters reveal clumpiness in star-forming regions and radio lobes. Importantly, non-parametric metrics do not require convergence of complex fitting algorithms, making them computationally efficient and robust even in low signal-to-noise regimes. Their application to radio continuum imaging provides systematic characterisation of source morphology without requiring assumptions about the underlying brightness profile, and can feed knowledge into ML pipelines for automated source classification and characterisation (see Section~\ref{sec:machine_learning}). 

\subsection{Parametric Decomposition and Multi-Component Characterisation}
\label{sec:parametric_decomposition}
Whilst morphological classification traditionally relies on visual inspection and phenomenological categorisation (e.g. FR~I/II, bent-tailed sources), parametric decomposition approaches offer a complementary pathway to source characterisation. These methods model the brightness distribution using mathematical functions, enabling quantitative measurements of structural parameters such as sizes, peak brightness, ellipticity, and concentration indices. Parametric decomposition has been widely employed in optical astronomy \citep{Simard_2002, Peng_2010} and more recently adapted for radio continuum studies \citep{loreto2015, Hodge2019, Song2021, Lucatelli_2024}.

The most commonly used parametric models in radio astronomy include Gaussian profiles \citep{Condon_1997,Condon_1998}, exponential functions, and the S\'ersic profile \citep{sersic1963, caon1993}. The S\'ersic law provides a generalised surface brightness distribution of the form
\begin{align}
	I(R) = I_n \exp\left\{-b_n\left[\left(\frac{R}{R_n}\right)^{1/n} - 1\right]\right\},
\end{align}
where $I_n$ is the intensity at the effective radius $R_n$ (enclosing half the total flux), $n$ is the S\'ersic index, and $b_n$ ensures that $R_n$ corresponds to the half-light radius. This law generalises both Gaussian ($n = 0.5$) and exponential ($n = 1$) profiles. Compact structures (e.g. AGN cores) typically follow Gaussian profiles resembling the synthesised beam, whilst extended emission often requires higher S\'ersic indices ($n \geq 1$).

\subsubsection{Multi-Component S\'ersic Fitting}
Radio sources frequently exhibit complex, multi-component structures spanning multiple spatial scales, including compact cores (AGN activity or starbursts), nuclear diffuse emission (circumnuclear star formation), and large-scale structures (galactic winds/outflows or extended star formation). Accurately characterising these components requires simultaneous fitting of multiple S\'ersic profiles to disentangle their contributions. The general model is a linear superposition
\begin{align}
	I_{\mathrm{total}}(x,y) = \sum_{i=1}^{N} I_i(x,y; \Theta_i),
\end{align}
where $I_i(x,y; \Theta_i)$ represents the $i$-th S\'ersic component with parameters $\Theta_i = \{I_{n,i},\allowbreak R_{n,i},\allowbreak x_{0,i},\allowbreak y_{0,i},\allowbreak q_i,\allowbreak \mathrm{PA}_i,\allowbreak n_i\}$. The model is fitted via non-linear least-squares minimisation using algorithms such as Levenberg-Marquardt \citep{lourakis04LM} or Trust Region Reflective methods \citep{Liu:2020:10.1016/j.compgeo.2020.103689, Branch:1996}.

However, multi-component fitting is highly sensitive to initial parameter estimates, and poor starting conditions can lead to non-physical solutions or convergence to local minima rather than the global optimum \citep{Haussler_2007, Andrae2011}. This is particularly problematic for sources with irregular morphologies, where the parameter space is large and degenerate, and small perturbations in initial conditions can produce large variations in best-fit parameters.

\subsubsection{Source Extraction as a Constraint for Parametric Fitting}
A critical advance in addressing the challenges of multi-component fitting lies in integrating source extraction algorithms to provide physically motivated initial conditions. Rather than relying on arbitrary parameter guesses, the source extraction tools described in Section~\ref{subsec:classical_source_finding} can identify distinct emission regions and compute basic photometric properties: peak brightness position $(x_0, y_0)$, half-light radius $R_{50}$, axis ratio $q = R_b / R_a$, and position angle (PA). Using these measured properties to initialise and constrain the S\'ersic fitting makes the minimisation process more efficient and robust, ensuring that fitted components correspond to actual structures rather than mathematical artefacts.

This method has been successfully applied to decompose U/LIRG radio emission into core-compact, nuclear extended, and large-scale diffuse components, enabling detailed measurements of flux densities, sizes, and brightness temperatures across multiple spatial scales \citep{Lucatelli_2024}. The technique generalises beyond radio astronomy; for instance, applying source extraction to optical or infrared images prior to S\'ersic fitting could improve bulge--disc decompositions in disturbed or merging systems.

Beyond classical source extraction algorithms, machine learning approaches (see Section~\ref{sec:machine_learning}) offer an alternative pathway for constraining parametric fits. Deep learning models trained to segment radio sources or predict structural parameters can provide data-driven priors that adapt to complex morphologies and varying noise conditions, potentially outperforming traditional methods in challenging regimes. This integration of classical algorithms, parametric fitting, and machine learning represents a promising direction for automated source characterisation in the SKAO era.

\subsubsection{Multi-Scale and Multi-Wavelength Decomposition}
Beyond multi-component fitting, decomposition techniques across spatial scales and frequency bands provide complementary approaches to characterising radio emission. Interferometric observations with different baseline configurations (e.g. \emph{e}-MERLIN and VLA, or MeerKAT and ASKAP) probe distinct spatial scales, with high-angular resolution images sensitive to compact structures and low-angular resolution images recovering diffuse emission. By combining observations at intermediate angular resolutions and multiple frequencies, it is possible to map the transition between nuclear and extended emission whilst disentangling different emission mechanisms.

The technical framework for multi-scale decomposition involves iterative comparison of images at different angular resolutions. Consider a high-resolution image $I_{\mathrm{hr}}$ (restoring beam $\Theta_{\mathrm{hr}}$) and a low-resolution image $I_{\mathrm{lr}}$ (restoring beam $\Theta_{\mathrm{lr}}$). To compare these, the high-resolution image is convolved to match the lower resolution:
\begin{align}
	I_{\mathrm{hr}}^{\mathrm{conv}} = I_{\mathrm{hr}} \ast \Theta_{\mathrm{conv}},
\end{align}
where $\Theta_{\mathrm{conv}}$ transforms $\Theta_{\mathrm{hr}}$ into $\Theta_{\mathrm{lr}}$. The residual emission
\begin{align}
	\mathcal{R}_{\mathrm{lr}} = I_{\mathrm{lr}} - f_\Theta \, I_{\mathrm{hr}}^{\mathrm{conv}},
\end{align}
where $f_\Theta = A(\Theta_{\mathrm{lr}})/A(\Theta_{\mathrm{hr}})$ accounts for beam area changes, isolates extended emission not captured at high resolution. Iterating over intermediate angular resolutions constructs a complete picture of emission distribution across spatial scales.

Extending this approach across multiple frequency bands enables construction of individual spectral energy distributions (SEDs) for each spatial component (Lucatelli et al. in preparation). Radio SEDs encode critical information about emission mechanisms: compact AGN cores often exhibit flat or inverted spectra due to synchrotron self-absorption, whilst extended star-forming regions display steep synchrotron spectra, and free-free thermal emission ($S_\nu \propto \nu^{-0.1}$) from H\,II can be significant in starbursts at higher frequencies due to recent ongoing SF activity. By performing consistent multi-component decomposition across all frequencies with the SKAO, it is possible to construct component-specific SEDs for compact cores, nuclear extended emission, and large-scale diffuse structures. This spatial-spectral approach yields simpler, more interpretable SEDs than traditional fitting methods that model the entire spectrum without spatial information, enabling robust estimates of thermal fractions, non-thermal spectral indices, and star formation rates on a component-by-component basis.

These multi-scale and multi-wavelength techniques have proven effective for characterising complex systems such as U/LIRGs. For SKAO, which will deliver data across unprecedented ranges of both angular scales (milliarcsecond VLBI to arcminute single-dish) and fractional bandwidths, such decomposition will be essential for disentangling AGN and star formation contributions and understanding feedback processes, accretion physics, and black hole-galaxy co-evolution.

\subsubsection{Synergies and Implications for SKAO Source Characterisation}
The integration of non-parametric metrics, source extraction, multi-component and multi-scale source decomposition, with multi-wavelength analysis represents a natural evolution towards more quantitative, robust, and automated source characterisation. These methods are well-suited to the SKAO era, where the volume and complexity of data will necessitate scalable pipelines capable of characterising billions of sources across diverse morphologies, spatial scales, and spectral behaviours.

Non-parametric (CASGM) and parametric approaches are complementary. Non-parametric quantities provide rapid, model-independent characterisation robust in low signal-to-noise regimes, ideal for initial classification and quality assessment. Parametric decomposition provides detailed physical measurements (sizes, flux densities, brightness temperatures), enabling quantitative comparison across sources and their individual components. Combined, these approaches allow SKAO pipelines to flag unusual morphologies for detailed follow-up whilst providing precise measurements for well-behaved sources.

Both metric types serve as valuable input features for machine learning classifiers. Concentration indices, asymmetry, Gini coefficients, S\'ersic indices, and effective radii, combined with multi-frequency spectral information, can train supervised classifiers distinguishing AGN, star-forming galaxies, and hybrid sources. The interpretability of these physically motivated features facilitates scientific understanding and model validation, bridging classical and data-driven approaches.

Multi-wavelength datasets provide a natural pathway for integrating classical methods with machine learning. Spectral indices, thermal fractions, and component-based SEDs carry direct physical information about emission mechanisms, enabling models to recognise patterns associated with specific processes (AGN feedback, starburst-driven winds, post-merger relaxation, etc.). ML can enhance multi-component fitting by predicting optimal initial conditions, whilst parametric fitting results can generate high-fidelity synthetic training datasets, linking the automation between classical and data-driven methods.

Computational scalability is critical for SKAO. Non-parametric metrics are inexpensive and suitable for near-real-time screening of billions of sources. Source extraction and parametric fitting are more demanding but can be parallelised efficiently across distributed computing infrastructures using scalable GPU optimisations such as \textsc{JaX} \citep{jax2018github}. Multi-scale and multi-wavelength decomposition can exploit hierarchical pipelines where coarse-scale decomposition informs finer-scale fitting, and low-frequency results constrain high-frequency models.

Polarization information provides an additional and highly diagnostic dimension for source characterisation. Polarized intensity, fractional polarization, depolarization behaviour, and Faraday rotation measures can help distinguish AGN-related synchrotron emission from star-forming processes, trace magnetic-field structures, and identify physical components that may not be evident from total-intensity morphology alone. Incorporating polarization products into future source characterisation pipelines will therefore be of particular value.

The quantitative structural and spectral parameters derived from these integrated methods will be essential for future SKAO scientific cases, including studies of AGN feedback, cosmic star formation history, and black hole-galaxy co-evolution. The SKAO, with its unparalleled sensitivity, angular resolution, and spectral coverage, will fully realise the potential of these characterisation techniques, providing complete catalogues of radio sources and their physical properties.

\subsection{Machine Learning in Classification}
\label{sec:machine_learning}
With the growing availability of large-scale radio survey data, machine learning techniques have evolved beyond simple classifiers toward more structured and scalable frameworks. These developments encompass both model-driven innovations and data-centric strategies that collectively enhance the accuracy, interpretability, and generalisation of automated classification systems. The following sections outline key advances in model architectures, data handling techniques, and learning paradigms that have shaped current approaches to radio galaxy classification.

\subsubsection{Model-Based Approaches}

In recent years, the rapid progress of artificial intelligence and deep learning has led to substantial innovation in model design and optimization. These developments have naturally extended into radio astronomy, a field increasingly characterized by large, high-dimensional, and data-driven analyses. Beyond incremental improvements to algorithmic performance, advances in model architectures have played a central role in achieving robust classification and morphological interpretation of radio sources.

\paragraph{CNN Architectures}
CNNs have become essential for automated radio galaxy morphology classification. The network architecture itself, its depth, filter organization, and feature hierarchy, largely govern the model's generalization and discriminative capabilities. Consequently, progressively deeper and more sophisticated architectures have been adapted from computer vision to astronomy, including AlexNet \citep{krizhevsky2012imagenet, Aniyan17}, VGG-16 \citep{ma_2019,wu2019claran}, and DenseNet \citep{Huang16,samudre2022data}. 

The optimal depth of a CNN varies by application. For example, \citet{Lukic2019b} explored four- and eight-layer networks (CONVNET4 and CONVNET8) for the LoTSS DR1 dataset, while \citet{Becker21} and \citet{tang2019} used architectures with up to thirteen layers. Their comparative studies showed that deeper architectures (e.g., CONVNET8) achieved slightly higher precision (94.3\%) than shallower networks or capsule models (89.7\%), likely due to the increased number of non-linear transformations and feature hierarchies. Overall, deeper models tend to learn more abstract representations, resulting in improved performance across diverse radio source morphologies.

\begin{figure}[htb!]
\begin{center}
\includegraphics[width=\columnwidth,height=7.5cm]{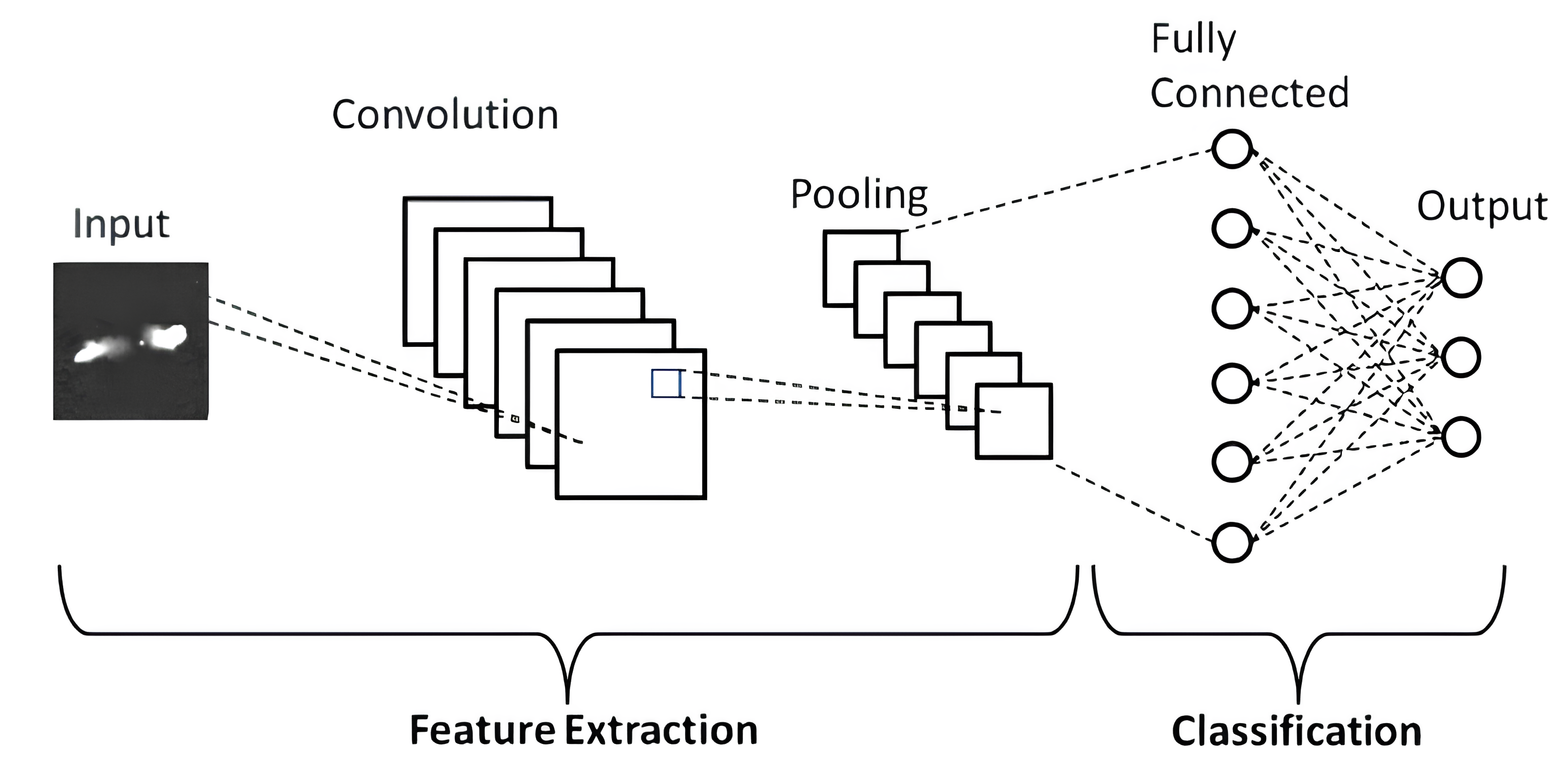}
\caption{Basic building blocks of a radio galaxy classifier based on a convolutional neural network (CNN).} 
\end{center}
\end{figure}

\paragraph{Regularization Techniques}

One of the major challenges in deep learning for radio astronomy is overfitting, primarily due to limited labeled data. To address this, various regularization methods are employed to improve model robustness and prevent memorization of the training set. 

A widely used approach is \textit{dropout}, which randomly deactivates a subset of neurons during training, thereby reducing co-adaptation of feature detectors \citep{tang2019, Tang22}. Another complementary method is \textit{batch normalization}, which standardizes feature map activations to stabilize and accelerate training by mitigating covariate shifts. These techniques collectively improve convergence speed and enhance generalization across heterogeneous survey data.

\paragraph{Specialized Convolutional Blocks}

Recent progress has also been driven by the introduction of specialized convolutional blocks and attention mechanisms designed to enhance model interpretability and performance.

Attention gates, inspired by human visual focus, enable models to emphasize salient image regions while suppressing irrelevant background features. \citet{Bowles21} incorporated attention mechanisms into CNN backbones, demonstrating a 50\% reduction in trainable parameters without compromising classification accuracy. The resulting attention maps also offer interpretable insights into model decision-making, advancing the goal of explainable AI in astronomy.

Another architectural innovation is the Group Equivariant CNN (G-CNN), which introduces rotational and reflectional equivariance into convolutional kernels \citep{cohen16}. This adaptation allows CNNs to maintain consistent feature representations under geometric transformations, a particularly valuable property given the arbitrary orientations of radio jets. When applied to the MiraBest dataset, G-CNNs demonstrated improved classification stability and accuracy \citep{Scaife21}.

Finally, multi-branch and multi-domain CNNs extend traditional models by incorporating multiple input modalities (e.g., total intensity, polarization maps, or multi-frequency data), enhancing the capacity to capture complex source structures \citep{Tang22}.

\subsubsection{Data-Centric Approaches}

While model architecture is critical, the quality and diversity of training data remain equally important. Radio astronomy data must be carefully curated, free from radio-frequency interference (RFI), imaging artefacts, or ambiguous labels to ensure meaningful learning outcomes. Given the scarcity of large, well-annotated datasets, data-centric strategies play a vital role in improving generalization.

\paragraph{Data Augmentation}

Data augmentation artificially increases dataset size and diversity by applying geometric and photometric transformations such as rotation, flipping, scaling, brightness adjustment, and contrast modification \citep{Aniyan17, Alhassan18, Lukic18, Becker21}. These techniques help mitigate class imbalance and improve rotational invariance, particularly for underrepresented morphologies like bent or compact sources. 

Advanced augmentation strategies use oversampling of minority classes or synthetic data generation with generative adversarial networks (GANs; \citealt{Rustige23}). However, care must be taken, as biases in the original data can be propagated or amplified through augmentation. Nonetheless, when properly designed, augmentation remains a cornerstone for robust training in limited-data regimes.

\paragraph{Rotation Invariance}

Since radio sources can appear at arbitrary orientations, models must exhibit rotational robustness. This can be achieved either through architectural design (e.g., G-CNNs) or through preprocessing techniques that standardize image orientation. Methods such as principal component analysis (PCA) can align source axes prior to model input, effectively normalizing rotation \citep{polsterer2019pink, Brand2023}. Combining augmentation and preprocessing has been shown to significantly enhance classification consistency across surveys.

\paragraph{Feature Engineering}

In classical machine learning pipelines, handcrafted features, derived from astrophysical or morphological insight, can provide compact, physically meaningful representations of data. Examples include peak brightness, lobe separation, texture measures (e.g., Haralick features; \citealt{ntwaetsile2021rapid}), and rotationally invariant descriptors based on Zernike moments \citep{sadeghi2021morphological}. Dimensionality reduction methods such as PCA are also used to compress features while retaining key variance \citep{darya2023morphological}. These engineered features are then input to algorithms such as random forests, gradient boosting, SVMs, or clustering methods like HDBSCAN, achieving classification accuracies exceeding 95\% in some cases. The main limitation, however, lies in the reliance on domain expertise and potential omission of subtle, high-dimensional information.

\subsubsection{Weak Supervision Approaches}

Given the high cost of manual annotation, weakly supervised paradigms are increasingly adopted to leverage vast unlabeled datasets. Three primary strategies have emerged in radio astronomy: transfer learning, self-supervised/semi-supervised learning, and few-shot learning.

\paragraph{Transfer Learning}

Transfer learning enables models pre-trained on large datasets to be fine-tuned on smaller, domain-specific samples. This approach effectively transfers learned representations from generic to specialized contexts, drastically reducing the required labeled data. Studies show that CNNs pre-trained on high-resolution surveys (e.g., FIRST; \citet{Becker95}) retain strong performance when fine-tuned on lower-resolution data such as NVSS \citep{tang2019}. Architectures like Inception-ResNet-v2 \citep{szegedy2017inception} and DenseNet have achieved accuracies above 90\% when applied to FR I/FR II classification \citep{Lukic2019b}.

\paragraph{Semi-Supervised and Self-Supervised Learning}

Self-supervised and semi-supervised learning (SSL) approaches have recently been explored to mitigate the scarcity of labeled data in radio astronomy. In self-supervised learning, models are first pretrained on unlabeled data using surrogate objectives, such as representation reconstruction or contrastive learning---and subsequently fine-tuned on limited annotated samples. Examples include autoencoder-based pretraining \citep{ma_2019}, as well as contrastive or masked-reconstruction frameworks applied to radio galaxy datasets \citep{hossain_2023, slijepcevic_2024, riggi_2024, cecconello_2024, lastufka_2024}.
In contrast, semi-supervised learning jointly leverages labeled and unlabeled data during training, typically through consistency regularization or pseudo-labeling techniques, as implemented in methods such as FixMatch \citep{Marti2022Dense}. Semi-supervised applications to radio source classification have shown promising results even with scarce annotations \citep{slijepcevic_2022, gupta_2023}.
Both self- and semi-supervised strategies have achieved high precision and recall (often exceeding 90\%) while substantially reducing the need for manual labeling, making them particularly relevant for SKAO-scale surveys, where unlabeled data vastly outnumbers labeled examples.

\paragraph{N-Shot Learning}

Few-shot and N-shot learning frameworks are designed for scenarios with minimal supervision. Siamese or prototypical networks learn similarity metrics rather than direct classification boundaries, enabling them to generalize from a few examples \citep{koch2015siamese, Samudre22}. Though performance remains somewhat lower than fully supervised CNNs, these methods represent a practical route toward scalable, label-efficient classification of radio sources.

\subsubsection{Label Ambiguity and Morphological Variability}
A significant challenge for supervised ML classification in radio astronomy is the inherent ambiguity of source labels. Galaxy morphologies do not fall into discrete, well-separated categories; instead, they form a continuum of structures that are often difficult to assign unambiguously to a single class. This ambiguity is further compounded by the strong dependence of observed morphology on frequency, angular resolution, and instrumental sensitivity. A source classified as compact or unresolved in one survey may exhibit extended or multi-component structure in another, while diffuse emission visible at low frequencies may be entirely absent at higher frequencies due to spectral index variations \citep[e.g.,][]{mb17, mingo19}.

These frequency- and resolution-dependent effects present a direct challenge for supervised classification models trained on data from a single survey. For example, a model trained on MeerKAT images may not generalise reliably to LOFAR or ASKAP data without retraining or domain adaptation, as differences in resolution, noise properties, and imaging characteristics can significantly alter source appearance. In addition, the use of human-assigned labels, whether from expert annotation or citizen science efforts such as Radio Galaxy Zoo \citep{banfield15}, introduces inter-annotator variability, particularly for ambiguous sources. This can propagate label noise into training datasets and degrade classifier performance.

Addressing these challenges requires careful consideration at both the data and model levels. Probabilistic or soft-label approaches, which assign a distribution over classes rather than a single deterministic label, can better capture intrinsic morphological uncertainty. Constructing multi-survey training datasets spanning a range of frequencies, resolutions, and sensitivities can further improve model generalisation, provided that label consistency is maintained across surveys. In addition, domain adaptation and transfer learning techniques offer promising pathways for adapting models trained on one survey to another without requiring complete retraining \citep{Aniyan17}. As the SKAO begins delivering data across diverse observing regimes, developing classification frameworks that are robust to such observational and labeling ambiguities will be essential for producing reliable, survey-independent source catalogs.

\section{Current Limitations and Challenges}
Despite significant progress in automated source detection and classification, several limitations continue to hinder the full exploitation of modern radio survey data. Traditional algorithms often struggle with complex, extended, or low-surface-brightness sources, leading to incomplete or fragmented detections. Conversely, ML and DL models, while more flexible, rely heavily on the quality, realism, and diversity of training datasets. The lack of comprehensive, labeled data that fully captures instrumental effects, calibration errors, and the diversity of astrophysical morphologies remains a major bottleneck.

To mitigate this dependency, emerging approaches such as self-supervised and active learning offer promising alternatives, enabling models to learn from partially labeled or even unlabeled data. These strategies can accelerate morphology discovery and adaptively refine models using human feedback where necessary.

Another key challenge lies in the generalization of trained models across different instruments and survey conditions. Models trained on specific datasets (e.g., LOFAR or ASKAP) often underperform when applied to data from other telescopes with differing resolutions, noise characteristics, or beam patterns. Domain adaptation techniques and standardized data representations are thus crucial to ensure robust transferability across observatories.

Most current frameworks are primarily morphology-based and do not yet exploit the spectral or temporal dimensions of radio data, which carry critical physical information about source evolution and emission mechanisms. Integrating spectral indices, variability, and polarization features into detection pipelines will be essential for a more complete astrophysical characterization.

From a computational standpoint, the petabyte-scale data expected from next-generation facilities such as the SKAO demands HPC- and cloud-ready architectures for scalable, real-time analysis. Efficient parallelization, data streaming, and memory optimization are key requirements, along with robust provenance tracking to ensure reproducibility across distributed infrastructures.

Finally, the next leap in automation requires joint source-finding and classification frameworks capable of performing end-to-end inference, from raw visibility data to labeled catalogs, without disjointed intermediate steps. Working directly in the Fourier domain, however, introduces additional challenges. The $uv$-plane representation of interferometric data is inherently non-uniform in sampling density, and the mapping between visibilities and sky brightness is non-trivial. Moreover, incomplete $uv$-coverage leads to correlated noise structures in the image plane. 

These factors imply that standard ML architectures, typically designed for regular image grids, cannot be directly applied to visibility data without modification. Instead, this requires either specialised network architectures capable of operating on irregular or graph-based representations, or carefully designed gridding and weighting schemes that preserve the statistical properties of the data \citep{Taran23, Drozdova24}. Community-wide benchmarking initiatives, such as the SKAO data challenges, remain vital for developing consistent performance metrics that balance completeness, reliability, and computational efficiency. Addressing these challenges will be essential for building next-generation, interpretable, and physically grounded radio-source analysis systems. Finally, while the current paradigm in radio astronomy has largely involved adapting ML methods developed in other fields, particularly computer vision and natural language processing, to astronomical data, there is growing potential for this relationship to become reciprocal. Radio survey data possess a number of statistically rich and scientifically distinctive properties: high-dimensional, multimodal feature spaces spanning morphology, spectral index, polarization, and variability; complex, correlated noise structures arising from interferometric sampling; and an enormous dynamic range in source properties across cosmic time. 

These characteristics present unique challenges that can drive the development of genuinely novel AI architectures and learning strategies, for example, methods for learning from irregular or sparse data representations, scalable uncertainty, aware inference, and self-supervised learning from physically structured data. As next-generation facilities such as the SKAO begin delivering data of unprecedented volume and complexity, the radio astronomy community is well positioned not only to benefit from advances in the broader AI field but also to contribute to it, with techniques and insights that may prove broadly applicable beyond astronomy.

\section{Future Directions}
The upcoming era of large-scale surveys with facilities such as the SKAO, ngVLA, and LOFAR 2.0 will demand a paradigm shift from isolated source detection tasks to fully integrated, intelligent data analysis platforms. These instruments will produce petabytes of data per day, revealing an unprecedented diversity of radio morphologies, spectral behaviors, and transient phenomena. To harness their full scientific potential, the community must advance toward scalable, interpretable, and physically informed machine learning frameworks.

A key direction is the development of multimodal learning systems that jointly analyze morphological, spectral, and temporal information. By fusing continuum maps with polarization, spectral index, and variability data, these models can capture the complete emission behavior of radio sources, enabling more physically meaningful classification. In particular, temporal-domain learning can facilitate the identification of episodic or recurrent AGN activity, while spectral–spatial networks may help disentangle overlapping sources in crowded fields.

In parallel, self-supervised, active, and transfer learning strategies will play a crucial role in reducing the dependence on manually labeled datasets. These approaches can leverage the vast quantities of unlabeled SKAO data, using generative or contrastive objectives to learn representations that generalize across instruments and observing conditions. Human-in-the-loop frameworks, where experts guide the model through targeted feedback, will further accelerate the discovery of new or rare morphological classes.

Future pipelines will likely evolve into end-to-end, cloud-native systems that integrate calibration, source finding, and classification within a unified workflow, deployable on HPC or distributed infrastructures for near-real-time analysis. Explainable AI (XAI) will be equally critical: interpretable models that highlight physically meaningful features—such as jet curvature, lobe symmetry, or spectral steepness- will build scientific trust and enable theory validation. In parallel, adherence to FAIR data principles will ensure interoperability across international facilities.

Ultimately, the synergy between next-generation surveys, physically motivated ML architectures, and collaborative open-science infrastructures will define the next decade of radio astronomy. This integrated approach will not only improve the completeness and reliability of radio-source catalogs but also open new pathways for uncovering the physical processes driving galaxy evolution across cosmic time.

\subsection{Integrated Detection and Classification Pipelines}
With the advent of large-scale sky surveys, the need for end-to-end automated pipelines that integrate source detection, characterization, and classification has become essential. Such integrated frameworks minimize human intervention and ensure consistent processing of large datasets while maintaining high accuracy and reproducibility. These pipelines combine multiple stages---calibration, source finding, morphological characterization, cross-identification, and classification---into a unified workflow.

Modern implementations exemplify this integration by coupling source detection algorithms with data quality assessment and catalog generation. Notable examples include the ASKAP pipeline \textsc{Selavy} \citep{Whiting2012}, the LOFAR \textsc{PyBDSF}-based framework \citep{Mohan2015}, and the MeerKAT \textsc{Oxkat} and \textsc{CARACal} systems \citep{2024IAUGA..32P1735R, 2020ASPC..527..635J}. These systems are often designed to operate within large-scale data processing environments such as the SKA Regional Centres (SRCs), supporting distributed computation and standardized metadata handling. The SRC network will play a central role in enabling these workflows by providing standardized execution environments, scalable computing resources, and access to common analysis services across the international SKAO community.

In the future, the combination of classical methods with machine-learning techniques is likely to become increasingly common in modern radio surveys. For example, \citet{Alegre22} used machine-learning techniques to assess the identification of host galaxies for LOFAR radio sources, while \citet{Silima25} tested whether supervised ML algorithms, including Random Forest and XGBoost, can reproduce source classifications into AGN and star-forming galaxies compared to the traditional diagnostic approach of \citet{Whittam22}. A hybrid approach is also planned for source association within the EMU survey \citep{Tang25}. Such hybrid architectures, in which rule-based methods address well-constrained tasks such as noise estimation and Gaussian component fitting while ML modules operate on higher-level inference and morphologically ambiguous cases, provide a practical and scalable template for SKAO-era pipelines, where the volume, complexity, and diversity of the detected source population will fundamentally exceed the capabilities of purely deterministic approaches.

\section{Implications for SKAO Science Goals}
The integration of advanced machine learning and automated source-finding pipelines has direct and far-reaching implications for the scientific objectives of the SKAO. As the SKAO aims to map billions of radio sources across cosmic time, the ability to detect, classify, and interpret these sources efficiently and accurately is central to its key science drivers, ranging from understanding galaxy evolution and cosmic magnetism to probing the cosmic dawn.

Accurate morphological classification of radio galaxies, including Fanaroff–Riley (FR) types, HyMoRSs, DDRGs, and GRGs, will enable a more complete census of AGN activity and feedback mechanisms. By tracing jet–environment interactions across redshift, SKAO datasets will shed light on how radio-mode feedback regulates star formation and black hole growth in galaxies. Automated pipelines that can identify these diverse morphologies at scale are therefore essential for linking radio structures to their host galaxy and environmental properties.

Furthermore, the improved recovery of diffuse and low-surface-brightness emission will directly benefit studies of cluster physics and large-scale structure formation. Enhanced sensitivity to bent-tailed sources and radio relics can provide new probes of intra-cluster medium dynamics, shock fronts, and magnetization processes in the cosmic web. Deep-learning–based extraction methods, capable of distinguishing faint extended features from noise, are thus crucial for advancing the SKAO's cosmic magnetism and large-scale structure science programs.

The SKAO will also explore transient and time-variable radio phenomena, ranging from episodic AGN activity to explosive events such as fast radio bursts (FRBs) and supernovae. Machine-learning frameworks that incorporate temporal and spectral dimensions will enable real-time detection and classification of these events, facilitating rapid multiwavelength follow-up and cross-correlation with optical, X-ray, and gravitational-wave observatories.

In summary, the advancement of automated, interpretable, and physically informed source detection and classification systems is not merely a technical enhancement but a scientific necessity for achieving the fundamental goals of the SKAO. By bridging data-driven methods with astrophysical insight and embedding these within open and interoperable infrastructures, such frameworks will enable the SKAO to move beyond simple source cataloging toward uncovering the physical mechanisms that govern the radio Universe.

\section{Conclusions}
The evolution of radio source finding and classification has mirrored the technological transformation of radio astronomy itself---from the early days of manual cataloging to the present era of data-intensive, automated, and intelligent pipelines. Classical algorithms, though foundational, were designed for datasets of limited scale and complexity. As surveys expanded in sensitivity and coverage through instruments such as LOFAR, MeerKAT, ASKAP, and the uGMRT, it became evident that traditional thresholding and Gaussian-fitting approaches could no longer meet the demands of modern data volumes or the diversity of radio source morphologies.

Machine learning and deep learning methods have emerged as effective tools to address these challenges, offering the ability to learn complex, hierarchical features directly from data. Architectures such as convolutional neural networks, U-Nets, and object-detection frameworks have demonstrated exceptional capability in identifying compact, extended, and diffuse emission structures across a wide range of signal-to-noise regimes. When integrated into scalable, HPC environments, these approaches promise to deliver real-time, reproducible, and scientifically reliable catalogues from the vast imaging data expected from the SKAO and its precursors.

The future lies in end-to-end, unified frameworks that jointly perform detection, classification, and characterization of sources, leveraging self-supervised learning, active learning, and transfer learning to adapt to heterogeneous data domains. Such systems must not only optimize completeness and reliability but also incorporate physical interpretability, ensuring that data-driven classifications retain astrophysical meaning.

Ultimately, the development of robust and intelligent source-finding pipelines is not merely a computational challenge but a scientific imperative. Accurate and scalable detection and classification underpin nearly every major SKA) science goal, from tracing the cosmic evolution of AGN and star formation to mapping magnetic fields and large-scale structure. The synergy between AI-driven methods, rigorous physical modeling, and high-performance computing will define the next generation of radio astronomy, unlocking the full scientific potential of the SKAO to probe the radio Universe.

\section*{Acknowledgements}
We thank Ivy Wong for her valuable insights on H\,\textsc{i} source-finding techniques. Omkar Bait was supported by the National Science Foundation under Cooperative Agreement 2421782 and the Simons Foundation grant MPS-AI-00010515 awarded to the NSF-Simons AI Institute for Cosmic Origins — CosmicAI, https://www.cosmicai.org/.  GL, AA, and JM acknowledge financial support from the Spanish grant PID2023-147883NB-C21, funded by MCIU/AEI/ 10.13039/501100011033, as well as support through ERDF/EU and from the Severo Ochoa grant CEX2021-001131-S funded by MCIN/AEI/ 10.13039/501100011033.

\bibliographystyle{abbrvnat-maxbibnames4}
\bibliography{chapter} 

@ARTICLE{Taran23,
       author = {{Taran}, O. and {Bait}, O. and {Dessauges-Zavadsky}, M. and {Holotyak}, T. and {Schaerer}, D. and {Voloshynovskiy}, S.},
        title = "{Challenging interferometric imaging: Machine learning-based source localization from uv-plane observations}",
      journal = {\aap},
         year = 2023,
       volume = {674},
          eid = {A161},
        pages = {A161},
          doi = {https://doi.org/10.1051/0004-6361/202245778},
archivePrefix = {arXiv},
       eprint = {2305.03533},
 primaryClass = {astro-ph.IM},
       adsurl = {https://ui.adsabs.harvard.edu/abs/2023A&A...674A.161T}
}

@ARTICLE{Drozdova24,
       author = {{Drozdova}, M. and {Kinakh}, V. and {Bait}, O. and {Taran}, O. and {Lastufka}, E. and {Dessauges-Zavadsky}, M. and {Holotyak}, T. and {Schaerer}, D. and {Voloshynovskiy}, S.},
        title = "{Radio-astronomical image reconstruction with a conditional denoising diffusion model}",
      journal = {\aap},
         year = 2024,
        
       volume = {683},
          eid = {A105},
        pages = {A105},
          doi = {https://doi.org/10.1051/0004-6361/202347948},
archivePrefix = {arXiv},
       eprint = {2402.10204},
 primaryClass = {astro-ph.IM},
       adsurl = {https://ui.adsabs.harvard.edu/abs/2024A&A...683A.105D}
}

@ARTICLE{hopkins15,
       author = {{Hopkins}, A.~M. and {Whiting}, M.~T. and {Seymour}, N. and {Chow}, K.~E. and {Norris}, R.~P. and {Bonavera}, L. and {Breton}, R. and {Carbone}, D. and {Ferrari}, C. and {Franzen}, T.~M.~O. and {Garsden}, H. and {Gonz{\'a}lez-Nuevo}, J. and {Hales}, C.~A. and {Hancock}, P.~J. and {Heald}, G. and {Herranz}, D. and {Huynh}, M. and {Jurek}, R.~J. and {L{\'o}pez-Caniego}, M. and {Massardi}, M. and {Mohan}, N. and {Molinari}, S. and {Orr{\`u}}, E. and {Paladino}, R. and {Pestalozzi}, M. and {Pizzo}, R. and {Rafferty}, D. and {R{\"o}ttgering}, H.~J.~A. and {Rudnick}, L. and {Schisano}, E. and {Shulevski}, A. and {Swinbank}, J. and {Taylor}, R. and {van der Horst}, A.~J.},
        title = "{The ASKAP/EMU Source Finding Data Challenge}",
      journal = {\pasa},
         year = 2015,
        
       volume = {32},
          eid = {e037},
        pages = {e037},
          doi = {https://doi.org/10.1017/pasa.2015.37},
archivePrefix = {arXiv},
       eprint = {1509.03931},
 primaryClass = {astro-ph.IM},
       adsurl = {https://ui.adsabs.harvard.edu/abs/2015PASA...32...37H}
}

@ARTICLE{Gheller2018,
       author = {{Gheller}, C. and {Vazza}, F. and {Bonafede}, A.},
        title = "{Deep learning based detection of cosmological diffuse radio sources}",
      journal = {\mnras},
         year = 2018,
        
       volume = {480},
       number = {3},
        pages = {3749-3761},
          doi = {https://doi.org/10.1093/mnras/sty2102},
archivePrefix = {arXiv},
       eprint = {1809.03315},
 primaryClass = {astro-ph.IM},
       adsurl = {https://ui.adsabs.harvard.edu/abs/2018MNRAS.480.3749G}
}

@ARTICLE{Vafaei2019,
       author = {{Vafaei Sadr}, A. and {Vos}, Etienne E. and {Bassett}, Bruce A. and {Hosenie}, Zafiirah and {Oozeer}, N. and {Lochner}, Michelle},
        title = "{DEEPSOURCE: point source detection using deep learning}",
      journal = {\mnras},
         year = 2019,
        
       volume = {484},
       number = {2},
        pages = {2793-2806},
          doi = {https://doi.org/10.1093/mnras/stz131},
archivePrefix = {arXiv},
       eprint = {1807.02701},
 primaryClass = {astro-ph.IM},
       adsurl = {https://ui.adsabs.harvard.edu/abs/2019MNRAS.484.2793V}
}

@ARTICLE{Lukic2019,
       author = {{Lukic}, Vesna and {de Gasperin}, Francesco and {Br{\"u}ggen}, Marcus},
        title = "{ConvoSource: Radio-Astronomical Source-Finding with Convolutional Neural Networks}",
      journal = {Galaxies},
         year = 2019,
        
       volume = {8},
       number = {1},
          eid = {3},
        pages = {3},
          doi = {https://doi.org/10.3390/galaxies8010003},
archivePrefix = {arXiv},
       eprint = {1910.03631},
 primaryClass = {astro-ph.IM},
       adsurl = {https://ui.adsabs.harvard.edu/abs/2019Galax...8....3L}
}

@inproceedings{he2017mask,
  title={Mask r-cnn},
  author={He, Kaiming and Gkioxari, Georgia and Doll{\'a}r, Piotr and Girshick, Ross},
  booktitle={Proceedings of the IEEE international conference on computer vision},
  pages={2961--2969},
  year={2017}
}

@inproceedings{ronneberger2015u,
  title={U-net: Convolutional networks for biomedical image segmentation},
  author={Ronneberger, Olaf and Fischer, Philipp and Brox, Thomas},
  booktitle={International Conference on Medical image computing and computer-assisted intervention},
  pages={234-241},
  year={2015},
  organization={Springer}
}

@inproceedings{pino2021semantic,
  title={Semantic segmentation of radio-astronomical images},
  author={Pino, Carmelo and Sortino, Renato and Sciacca, Eva and Riggi, Simone and Spampinato, Concetto},
  booktitle={International Workshop on Artificial Intelligence and Pattern Recognition},
  pages={393-403},
  year={2021},
  organization={Springer}
}

@ARTICLE{Hancock2012,
       author = {{Hancock}, P.~J. and {Murphy}, T. and {Gaensler}, B.~M. and {Hopkins}, A. and {Curran}, J.~R.},
        title = "{Compact continuum source finding for next generation radio surveys}",
      journal = {\mnras},
         year = 2012,
        
       volume = {422},
       number = {2},
        pages = {1812-1824},
          doi = {https://doi.org/10.1111/j.1365-2966.2012.20768.x},
archivePrefix = {arXiv},
       eprint = {1202.4500},
 primaryClass = {astro-ph.IM},
       adsurl = {https://ui.adsabs.harvard.edu/abs/2012MNRAS.422.1812H}
}

@ARTICLE{Hancock2018,
       author = {{Hancock}, Paul J. and {Trott}, Cathryn M. and {Hurley-Walker}, Natasha},
        title = "{Source Finding in the Era of the SKA (Precursors): Aegean 2.0}",
      journal = {\pasa},
         year = 2018,
        
       volume = {35},
          eid = {e011},
        pages = {e011},
          doi = {https://doi.org/10.1017/pasa.2018.3},
archivePrefix = {arXiv},
       eprint = {1801.05548},
 primaryClass = {astro-ph.IM},
       adsurl = {https://ui.adsabs.harvard.edu/abs/2018PASA...35...11H}
}

@software{Mohan2015,
       author = {{Mohan}, Niruj and {Rafferty}, David},
        title = "{PyBDSF: Python Blob Detection and Source Finder}",
 howpublished = {Astrophysics Source Code Library, record ascl:1502.007},
         year = 2015,
        
          eid = {ascl:1502.007},
archivePrefix = {ascl},
       eprint = {1502.007},
       adsurl = {https://ui.adsabs.harvard.edu/abs/2015ascl.soft02007M}
}

@ARTICLE{Serra2015,
       author = {{Serra}, Paolo and {Westmeier}, Tobias and {Giese}, Nadine and {Jurek}, Russell and {Fl{\"o}er}, Lars and {Popping}, Attila and {Winkel}, Benjamin and {van der Hulst}, Thijs and {Meyer}, Martin and {Koribalski}, B{\"a}rbel S. and {Staveley-Smith}, Lister and {Courtois}, H{\'e}l{\`e}ne},
        title = "{SOFIA: a flexible source finder for 3D spectral line data}",
      journal = {\mnras},
         year = 2015,
        
       volume = {448},
       number = {2},
        pages = {1922-1929},
          doi = {https://doi.org/10.1093/mnras/stv079},
archivePrefix = {arXiv},
       eprint = {1501.03906},
 primaryClass = {astro-ph.IM},
       adsurl = {https://ui.adsabs.harvard.edu/abs/2015MNRAS.448.1922S}
}

@ARTICLE{sofia2,
       author = {{Westmeier}, T. and {Kitaeff}, S. and {Pallot}, D. and {Serra}, P. and {van der Hulst}, J.~M. and {Jurek}, R.~J. and {Elagali}, A. and {For}, B.-Q. and {Kleiner}, D. and {Koribalski}, B.~S. and {Lee-Waddell}, K. and {Mould}, J.~R. and {Reynolds}, T.~N. and {Rhee}, J. and {Staveley-Smith}, L.},
        title = "{SOFIA 2 - an automated, parallel H I source finding pipeline for the WALLABY survey}",
      journal = {\mnras},
         year = 2021,
        
       volume = {506},
       number = {3},
        pages = {3962-3976},
          doi = {https://doi.org/10.1093/mnras/stab1881},
archivePrefix = {arXiv},
       eprint = {2106.15789},
 primaryClass = {astro-ph.IM},
       adsurl = {https://ui.adsabs.harvard.edu/abs/2021MNRAS.506.3962W}
}

@ARTICLE{Whiting2012,
       author = {{Whiting}, Matthew T.},
        title = "{DUCHAMP: a 3D source finder for spectral-line data}",
      journal = {\mnras},
         year = 2012,
        
       volume = {421},
       number = {4},
        pages = {3242-3256},
          doi = {10.1111/j.1365-2966.2012.20548.x},
archivePrefix = {arXiv},
       eprint = {1201.2710},
 primaryClass = {astro-ph.IM},
       adsurl = {https://ui.adsabs.harvard.edu/abs/2012MNRAS.421.3242W}
}

@article{makovoz_2005,
  title={Compact continuum source finding for next generation radio surveys},
  author={Makovoz, David and Marleau, Francine R.},
  journal={Publications of the Astronomical Society of the Pacific
},
  volume={117},
  pages={1113},
  year={2005},
  doi={https://doi.org/10.1086/432977}
}

@INPROCEEDINGS{YOLO,
  author={Redmon, Joseph and Divvala, Santosh and Girshick, Ross and Farhadi, Ali},
  booktitle={2016 IEEE Conference on Computer Vision and Pattern Recognition (CVPR)}, 
  title={You Only Look Once: Unified, Real-Time Object Detection}, 
  year={2016},
  volume={},
  number={},
  pages={779-788},
  doi={https://doi.org/10.1109/CVPR.2016.91}
}

@ARTICLE{Burke2019,
       author = {{Burke}, Colin J. and {Aleo}, Patrick D. and {Chen}, Yu-Ching and {Liu}, Xin and {Peterson}, John R. and {Sembroski}, Glenn H. and {Lin}, Joshua Yao-Yu},
        title = "{Deblending and classifying astronomical sources with Mask R-CNN deep learning}",
      journal = {\mnras},
         year = 2019,
        
       volume = {490},
       number = {3},
        pages = {3952-3965},
          doi = {https://doi.org/10.1093/mnras/stz2845},
archivePrefix = {arXiv},
       eprint = {1908.02748},
 primaryClass = {astro-ph.IM},
       adsurl = {https://ui.adsabs.harvard.edu/abs/2019MNRAS.490.3952B}
}

@ARTICLE{Lao2023,
       author = {{Lao}, B. and {Jaiswal}, S. and {Zhao}, Z. and {Lin}, L. and {Wang}, J. and {Sun}, X. and {Qin}, S. -L.},
        title = "{Radio sources segmentation and classification with deep learning}",
      journal = {Astronomy and Computing},
         year = 2023,
        
       volume = {44},
          eid = {100728},
        pages = {100728},
          doi = {https://doi.org/10.1016/j.ascom.2023.100728},
archivePrefix = {arXiv},
       eprint = {2306.01426},
 primaryClass = {astro-ph.IM},
       adsurl = {https://ui.adsabs.harvard.edu/abs/2023A&C....4400728L}
}

@ARTICLE{Riggi2023,
       author = {{Riggi}, S. and {Magro}, D. and {Sortino}, R. and {De Marco}, A. and {Bordiu}, C. and {Cecconello}, T. and {Hopkins}, A.~M. and {Marvil}, J. and {Umana}, G. and {Sciacca}, E. and {Vitello}, F. and {Bufano}, F. and {Ingallinera}, A. and {Fiameni}, G. and {Spampinato}, C. and {Zarb Adami}, K.},
        title = "{Astronomical source detection in radio continuum maps with deep neural networks}",
      journal = {Astronomy and Computing},
         year = 2023,
        
       volume = {42},
          eid = {100682},
        pages = {100682},
          doi = {https://doi.org/10.1016/j.ascom.2022.100682},
archivePrefix = {arXiv},
       eprint = {2212.02538},
 primaryClass = {astro-ph.IM},
       adsurl = {https://ui.adsabs.harvard.edu/abs/2023A&C....4200682R}
}

@ARTICLE{Gupta2024,
       author = {{Gupta}, Nikhel and {Hayder}, Zeeshan and {Norris}, Ray P. and {Huynh}, Minh and {Petersson}, Lars},
        title = "{RadioGalaxyNET: Dataset and novel computer vision algorithms for the detection of extended radio galaxies and infrared hosts}",
      journal = {\pasa},
         year = 2024,
        
       volume = {41},
          eid = {e001},
        pages = {e001},
          doi = {https://doi.org/10.1017/pasa.2023.64},
archivePrefix = {arXiv},
       eprint = {2312.00306},
 primaryClass = {astro-ph.IM},
       adsurl = {https://ui.adsabs.harvard.edu/abs/2024PASA...41....1G}
}

@ARTICLE{Cornu2024,
       author = {{Cornu}, D. and {Salom{\'e}}, P. and {Semelin}, B. and {Marchal}, A. and {Freundlich}, J. and {Aicardi}, S. and {Lu}, X. and {Sainton}, G. and {Mertens}, F. and {Combes}, F. and {Tasse}, C.},
        title = "{YOLO-CIANNA: Galaxy detection with deep learning in radio data: I. A new YOLO-inspired source detection method applied to the SKAO SDC1}",
      journal = {\aap},
         year = 2024,
        
       volume = {690},
          eid = {A211},
        pages = {A211},
          doi = {https://doi.org/10.1051/0004-6361/202449548},
archivePrefix = {arXiv},
       eprint = {2402.05925},
 primaryClass = {astro-ph.IM},
       adsurl = {https://ui.adsabs.harvard.edu/abs/2024A&A...690A.211C}
}

@article{ntwaetsile2021rapid,
  title={Rapid sorting of radio galaxy morphology using Haralick features},
  author={Ntwaetsile, Kushatha and Geach, James E},
  journal={Monthly Notices of the Royal Astronomical Society},
  volume={502},
  number={3},
  pages={3417-3425},
  doi = {https://doi.org/10.1093/mnras/stab271},
  year={2021},
  publisher={Oxford University Press}
}

@article{sadeghi2021morphological,
  title={Morphological-based classifications of radio galaxies using supervised machine-learning methods associated with image moments},
  author={Sadeghi, Mohammad and Javaherian, Mohsen and Miraghaei, Halime},
  journal={The Astronomical Journal},
  volume={161},
  number={2},
  pages={94},
  year={2021},
  doi = {https://doi.org/10.3847/1538-3881/abd314},
  publisher={IOP Publishing}
}

@ARTICLE{riggi_2016,
       author = {{Riggi}, S. and {Ingallinera}, A. and {Leto}, P. and {Cavallaro}, F. and {Bufano}, F. and {Schillir{\`o}}, F. and {Trigilio}, C. and {Umana}, G. and {Buemi}, C.~S. and {Norris}, R.~P.},
        title = "{Automated detection of extended sources in radio maps: progress from the SCORPIO survey}",
      journal = {\mnras},
         year = 2016,
       volume = {460},
       number = {2},
        pages = {1486-1499},

          doi = {https://doi.org/10.1093/mnras/stw982},

}

@ARTICLE{riggi_2019,
       author = {{Riggi}, S. and {Vitello}, F. and {Becciani}, U. and {Buemi}, C. and {Bufano}, F. and {Calanducci}, A. and {Cavallaro}, F. and {Costa}, A. and {Ingallinera}, A. and {Leto}, P. and {Loru}, S. and {Norris}, R.~P. and {Schillir{\`o}}, F. and {Sciacca}, E. and {Trigilio}, C. and {Umana}, G.},
        title = "{Cuc(aesar) source finder: Recent developments and testing}",
      journal = {\pasa},
         year = 2019,
        
       volume = {36},
          eid = {e037},
        pages = {e037},
          doi = {https://doi.org/10.1017/pasa.2019.29},
archivePrefix = {arXiv},
       eprint = {1909.06116},
 primaryClass = {astro-ph.IM},
       adsurl = {https://ui.adsabs.harvard.edu/abs/2019PASA...36...37R}
}

@ARTICLE{robotham_2018,
       author = {{Robotham}, A.~S.~G. and {Davies}, L.~J.~M. and {Driver}, S.~P. and {Koushan}, S. and {Taranu}, D.~S. and {Casura}, S. and {Liske}, J.},
        title = "{ProFound: Source Extraction and Application to Modern Survey Data}",
      journal = {\mnras},
         year = 2018,
        
       volume = {476},
       number = {3},
        pages = {3137-3159},
          doi = {https://doi.org/10.1093/mnras/sty440},
archivePrefix = {arXiv},
       eprint = {1802.00937},
 primaryClass = {astro-ph.IM},
       adsurl = {https://ui.adsabs.harvard.edu/abs/2018MNRAS.476.3137R}
}

@ARTICLE{hale_2019,
       author = {{Hale}, C.~L. and {Robotham}, A.~S.~G. and {Davies}, L.~J.~M. and {Jarvis}, M.~J. and {Driver}, S.~P. and {Heywood}, I.},
        title = "{Radio source extraction with PROFOUND}",
      journal = {\mnras},
         year = 2019,
        
       volume = {487},
       number = {3},
        pages = {3971-3989},
          doi = {https://doi.org/10.1093/mnras/stz1462},
archivePrefix = {arXiv},
       eprint = {1902.01440},
 primaryClass = {astro-ph.GA},
       adsurl = {https://ui.adsabs.harvard.edu/abs/2019MNRAS.487.3971H}
}

@ARTICLE{carbone_2018,
       author = {{Carbone}, D. and {Garsden}, H. and {Spreeuw}, H. and {Swinbank}, J.~D. and {van der Horst}, A.~J. and {Rowlinson}, A. and {Broderick}, J.~W. and {Rol}, E. and {Law}, C. and {Molenaar}, G. and {Wijers}, R.~A.~M.~J.},
        title = "{PySE: Software for extracting sources from radio images}",
      journal = {Astronomy and Computing},
         year = 2018,
        
       volume = {23},
          eid = {92},
        pages = {92},
          doi = {https://doi.org/10.1016/j.ascom.2018.02.003},
archivePrefix = {arXiv},
       eprint = {1802.09604},
 primaryClass = {astro-ph.IM},
       adsurl = {https://ui.adsabs.harvard.edu/abs/2018A&C....23...92C}
}

@ARTICLE{hales_2012,
       author = {{Hales}, C.~A. and {Murphy}, T. and {Curran}, J.~R. and {Middelberg}, E. and {Gaensler}, B.~M. and {Norris}, R.~P.},
        title = "{BLOBCAT: software to catalogue flood-filled blobs in radio images of total intensity and linear polarization}",
      journal = {\mnras},
         year = 2012,
        
       volume = {425},
       number = {2},
        pages = {979-996},
          doi = {https://doi.org/10.1111/j.1365-2966.2012.21373.x},
archivePrefix = {arXiv},
       eprint = {1205.5313},
 primaryClass = {astro-ph.IM},
       adsurl = {https://ui.adsabs.harvard.edu/abs/2012MNRAS.425..979H}
}

@article{lopez_2012,
author={López-Caniego, M. and Vielva, P.},
title={Biparametric adaptive filter: detection of compact sources in complex microwave backgrounds},
journal={\mnras},
year={2012},
volume={421},
pages={2139},
doi={https://doi.org/10.1111/j.1365-2966.2012.20444.x}
}

@article{lopez_2006,
author={López-Caniego, M. and Herranz, D. and González-Nuevo, J. and Sanz, J. L. and Barreiro, R. B. and Vielva, P. and Argüeso, F. and Toffolatti, L.},
title={Comparison of filters for the detection of point sources in Planck simulations},
journal={\mnras},
year={2006},
volume={370},
pages={2047},
doi={https://doi.org/10.1111/j.1365-2966.2006.10639.x}
}

@ARTICLE{white_97,
       author = {{White}, Richard L. and {Becker}, Robert H. and {Helfand}, David J. and {Gregg}, Michael D.},
        title = "{A Catalog of 1.4 GHz Radio Sources from the FIRST Survey}",
      journal = {\apj},
         year = 1997,
        
       volume = {475},
       number = {2},
        pages = {479},
          doi = {https://doi.org/10.1086/303564},
       adsurl = {https://ui.adsabs.harvard.edu/abs/1997ApJ...475..479W}
}

@article{whiting_2012,
  title={Source-Finding for the Australian Square Kilometre Array Pathfinder},
  author={Whiting, M. and Humphreys, B.},
  journal={\pasa},
  volume={29},
  pages={371},
  year={2012},
  doi = {https://doi.org/10.1071/AS12028}
}

@article{bertin_1996,
author={Bertin, E. and Arnouts, S.},
title={SExtractor: Software for source extraction},
journal={Astronomy \& Astrophysics Suppllement Series},
year={1996},
volume={117},
pages={393},
doi={https://doi.org/10.1051/aas:1996164}
}

@article{spreeuw_2010,
  title={Search and detection of low frequency radio transients},
  author={Spreeuw, Johannes Norbertus},
  journal={Dissertation, Astronomical Institute Anton Pannekoek, University of Amsterdam},
  volume={ISBN 978-90-9024055-8},
  year={2010}
}

@article{swinbank_2015,
    author = {Swinbank, John D. and Staley, Tim D. and Molenaar, Gijs J. and Rol, Evert and Rowlinson, Antonia and Scheers, Bart and Spreeuw, Hanno and Bell, Martin E. and Broderick, Jess W. and Carbone, Dario and Garsden, Hugh and van der Horst, Alexander J. and Law, Casey J. and Wise, Michael and Breton, Rene P. and Cendes, Yvette and Corbel, Stéphane and Eislöffel, Jochen and Falcke, Heino and Fender, Rob Grießmeier, Jean-Mathias and Hessels, Jason W. T. and Stappers, Benjamin W. and Stewart, Adam J. and Wijers, Ralph A. M. J. and Wijnands, Rudy and Zarka, Philippe},
    title={The LOFAR Transients Pipeline},
    journal={Astronomy and Computing},
    volume={11},
    pages={25},
    year={2015},
    doi={https://doi.org/10.1016/j.ascom.2015.03.002}
}

@ARTICLE{koribalski_2020,
       author = {{Koribalski}, B{\"a}rbel S. and {Staveley-Smith}, L. and {Westmeier}, T. and {Serra}, P. and {Spekkens}, K. and {Wong}, O.~I. and {Lee-Waddell}, K. and {Lagos}, C.~D.~P. and {Obreschkow}, D. and {Ryan-Weber}, E.~V. and {Zwaan}, M. and {Kilborn}, V. and {Bekiaris}, G. and {Bekki}, K. and {Bigiel}, F. and {Boselli}, A. and {Bosma}, A. and {Catinella}, B. and {Chauhan}, G. and {Cluver}, M.~E. and {Colless}, M. and {Courtois}, H.~M. and {Crain}, R.~A. and {de Blok}, W.~J.~G. and {D{\'e}nes}, H. and {Duffy}, A.~R. and {Elagali}, A. and {Fluke}, C.~J. and {For}, B.-Q. and {Heald}, G. and {Henning}, P.~A. and {Hess}, K.~M. and {Holwerda}, B.~W. and {Howlett}, C. and {Jarrett}, T. and {Jones}, D.~H. and {Jones}, M.~G. and {J{\'o}zsa}, G.~I.~G. and {Jurek}, R. and {J{\"u}tte}, E. and {Kamphuis}, P. and {Karachentsev}, I. and {Kerp}, J. and {Kleiner}, D. and {Kraan-Korteweg}, R.~C. and {L{\'o}pez-S{\'a}nchez}, {\'A}. R. and {Madrid}, J. and {Meyer}, M. and {Mould}, J. and {Murugeshan}, C. and {Norris}, R.~P. and {Oh}, S.-H. and {Oosterloo}, T.~A. and {Popping}, A. and {Putman}, M. and {Reynolds}, T.~N. and {Rhee}, J. and {Robotham}, A.~S.~G. and {Ryder}, S. and {Schr{\"o}der}, A.~C. and {Shao}, Li and {Stevens}, A.~R.~H. and {Taylor}, E.~N. and {van{\^A} der Hulst}, J.~M. and {Verdes-Montenegro}, L. and {Wakker}, B.~P. and {Wang}, J. and {Whiting}, M. and {Winkel}, B. and {Wolf}, C.},
        title = "{WALLABY {\textendash} an SKA Pathfinder H I survey}",
      journal = {\apss},
         year = 2020,
        
       volume = {365},
       number = {7},
          eid = {118},
        pages = {118},
          doi = {https://doi.org/10.1007/s10509-020-03831-4},
archivePrefix = {arXiv},
       eprint = {2002.07311},
 primaryClass = {astro-ph.GA},
       adsurl = {https://ui.adsabs.harvard.edu/abs/2020Ap&SS.365..118K}
}

@article{amic_2011,
  title={10C survey of radio sources at 15.7 GHz - I. Observing, mapping and source extraction},
  author={AMI Consortium: Franzen, Thomas M. O. and Davies, Matthew L. and Waldram, Elizabeth M. and Grainge, Keith J. B. and Hobson, Michael P. and Hurley-Walker, Natasha and Lasenby, Anthony and Olamaie, Malak and Pooley, Guy G. and Rodríguez-Gonzálvez, Carmen and Saunders, Richard D. E. and Scaife, Anna M. M. and Schammel, Michel P. and Scott, Paul F. and Shimwell, Timothy W. and Titterington, David J. and Zwart, Jonathan T. L.},
  journal={\mnras},
  volume={415},
  pages={2699},
  year={2011},
  doi = {https://doi.org/10.1111/j.1365-2966.2011.18887.x}
}

@article{bonaldi_2021,
  title={Square Kilometre Array Science Data Challenge 1: analysis and results},
  author={Bonaldi, A. and An, T. and Brüggen, M. and Burkutean, S. and Coelho, B. and Goodarzi, H. and Hartley, P. and Sandhu, P. K. and Wu, C. and Yu, L. and Zhoolideh Haghighi, M. H. and Antón, S. and Bagheri, Z. and Barbosa, D. and Barraca, J. P. and Bartashevich, D. and Bergano, M. and Bonato, M. and Brand, J. and de Gasperin, F. Giannetti, A. and Dodson, R. and Jain, P. and Jaiswal, S. and Lao, B. and Liu, B. and Liuzzo, E. and Lu, Y. and Lukic, V. and Maia, D. and Marchili, N. and Massardi, M. and Mohan, P. and Morgado, J. B. and Panwar, M. and Prabhakar, P. and Ribeiro, V. A. R. M. and Rygl, K. L. J. and Sabz Ali, V. and Saremi, E. and Schisano, E. and Sheikhnezami, S. and Vafaei Sadr, A. and Wong, A. and Wong, O. I.},
  journal={\mnras},
  pages={3821},
  volume={500},
  year={2021},
  doi = {https://doi.org/10.1093/mnras/staa3023}
}

@article{fender_2007,
  title={LOFAR Transients and the Radio Sky Monitor},
  author={Fender, Rob and Wijers, Ralph, Ben and Stappers, Ban and Braun, Robert and Wise, Michael and Coenen, Thijs and Falcke, Heino and Griessmeier, Jean-Mathias and van Haarlem, Michiel and de Bruyn, Ger and Jonker, Peter and Law, Casey and Markoff, Sera and Masters, Joseph and Miller-Jones, James and Osten, Rachel and Scheers, Bart and Spreeuw, Hanno and Swinbank, John and Vogt, Corina Wijnandsb, Rudy and Zarka, Philippe},
  journal={Proceedings of Science},
  year={2007},
  doi={https://doi.org/10.48550/arXiv.0805.4349}
}

@article{haarlem_2013,
  title={LOFAR: The LOw-Frequency ARray},
  author={van Haarlem, M. P. and Wise, M. W. and Gunst, A. W. and Heald, G. and McKean, J. P. and Hessels, J. W. T. and de Bruyn, A. G. and Nijboer, R. and Swinbank, J. and Fallows, R. and Brentjens, M. and Nelles, A. and Beck, R. and Falcke, H. and Fender, R. and Hörandel, J. and Koopmans, L. V. E. and Mann, G. and Miley, G. and Röttgering, H. Stappers, B. W. and Wijers, R. A. M. J. and Zaroubi, S. and van den Akker, M. and Alexov, A. and Anderson, J. and Anderson, K. and van Ardenne, A. and Arts, M. and Asgekar, A. and Avruch, I. M. and Batejat, F. and Bähren, L. and Bell, M. E. and Bell, M. R. and van Bemmel, I. and Bennema, P. and Bentum, M. J. and Bernardi, G. and Best, P. and Bîrzan, L. and Bonafede, A. and Boonstra, A. -J. and Braun, R. and Bregman, J. and Breitling, F. and van de Brink, R. H. and Broderick, J. and Broekema, P. C. and Brouw, W. N. and Brüggen, M. and Butcher, H. R. and van Cappellen, W. and Ciardi, B. and Coenen, T. and Conway, J. and Coolen, A. and Corstanje, A. and Damstra, S. and Davies, O. and Deller, A. T. and Dettmar, R. -J. and van Diepen, G. and Dijkstra, K. and Donker, P. and Doorduin, A. and Dromer, J. and Drost, M. and van Duin, A. and Eislöffel, J. and van Enst, J. and Ferrari, C. and Frieswijk, W. and Gankema, H. and Garrett, M. A. and de Gasperin, F. and Gerbers, M. and de Geus, E. and Grießmeier, J. -M. and Grit, T. and Gruppen, P. and Hamaker, J. P. and Hassall, T. and Hoeft, M. and Holties, H. A. and Horneffer, A. and van der Horst, A. and van Houwelingen, A. and Huijgen, A. and Iacobelli, M. and Intema, H. and Jackson, N. and Jelic, V. and de Jong, A. and Juette, E. and Kant, D. and Karastergiou, A. and Koers, A. and Kollen, H. and Kondratiev, V. I. and Kooistra, E. and Koopman, Y. and Koster, A. and Kuniyoshi, M. and Kramer, M. and Kuper, G. and Lambropoulos, P. and Law, C. and van Leeuwen, J. and Lemaitre, J. and Loose, M. and Maat, P. and Macario, G. and Markoff, S. and Masters, J. and McFadden, R. A. and McKay-Bukowski, D. and Meijering, H. and Meulman, H. and Mevius, M. and Middelberg, E. and Millenaar, R. and Miller-Jones, J. C. A. and Mohan, R. N. and Mol, J. D. and Morawietz, J. and Morganti, R. and Mulcahy, D. D. and Mulder, E. and Munk, H. and Nieuwenhuis, L. and van Nieuwpoort, R. and Noordam, J. E. and Norden, M. and Noutsos, A. and Offringa, A. R. and Olofsson, H. and Omar, A. and Orrú, E. and Overeem, R. and Paas, H. and Pandey-Pommier, M. and Pandey, V. N. and Pizzo, R. and Polatidis, A. and Rafferty, D. and Rawlings, S. and Reich, W. and de Reijer, J. -P. and Reitsma, J. and Renting, G. A. and Riemers, P. and Rol, E. and Romein, J. W. and Roosjen, J. and Ruiter, M. and Scaife, A. and van der Schaaf, K. and Scheers, B. and Schellart, P. and Schoenmakers, A. and Schoonderbeek, G. and Serylak, M. and Shulevski, A. and Sluman, J. and Smirnov, O. and Sobey, C. and Spreeuw, H. and Steinmetz, M. and Sterks, C. G. M. and Stiepel, H. -J. and Stuurwold, K. and Tagger, M. and Tang, Y. and Tasse, C. and Thomas, I. and Thoudam, S. and Toribio, M. C. and van der Tol, B. and Usov, O. and van Veelen, M. and van der Veen, A. -J. and ter Veen, S. and Verbiest, J. P. W. and Vermeulen, R. and Vermaas, N. and Vocks, C. and Vogt, C. and de Vos, M. and van der Wal, E. and van Weeren, R. and Weggemans, H. and Weltevrede, P. and White, S. and Wijnholds, S. J. and Wilhelmsson, T. and Wucknitz, O. and Yatawatta, S. and Zarka, P. and Zensus, A. and van Zwieten, J.},
  journal={\aap},
  volume={556},
  year={2013},
  pages={A2},
  doi={https://doi.org/10.1051/0004-6361/201220873}
}

@ARTICLE{banfield15,
       author = {{Banfield}, J.~K. and {Wong}, O.~I. and {Willett}, K.~W. and {Norris}, R.~P. and {Rudnick}, L. and {Shabala}, S.~S. and {Simmons}, B.~D. and {Snyder}, C. and {Garon}, A. and {Seymour}, N. and {Middelberg}, E. and {Andernach}, H. and {Lintott}, C.~J. and {Jacob}, K. and {Kapi{\'n}ska}, A.~D. and {Mao}, M.~Y. and {Masters}, K.~L. and {Jarvis}, M.~J. and {Schawinski}, K. and {Paget}, E. and {Simpson}, R. and {Kl{\"o}ckner}, H. -R. and {Bamford}, S. and {Burchell}, T. and {Chow}, K.~E. and {Cotter}, G. and {Fortson}, L. and {Heywood}, I. and {Jones}, T.~W. and {Kaviraj}, S. and {L{\'o}pez-S{\'a}nchez}, {\'A}. R. and {Maksym}, W.~P. and {Polsterer}, K. and {Borden}, K. and {Hollow}, R.~P. and {Whyte}, L.},
        title = "{Radio Galaxy Zoo: host galaxies and radio morphologies derived from visual inspection}",
      journal = {\mnras},
         year = 2015,
        
       volume = {453},
       number = {3},
        pages = {2326},
          doi = {https://doi.org/10.1093/mnras/stv1688},
archivePrefix = {arXiv},
       eprint = {1507.07272},
 primaryClass = {astro-ph.GA},
       adsurl = {https://ui.adsabs.harvard.edu/abs/2015MNRAS.453.2326B}
}

@ARTICLE{Brand2023,
       author = {{Brand}, Kevin and {Grobler}, Trienko L. and {Kleynhans}, Waldo and {Vaccari}, Mattia and {Prescott}, Matthew and {Becker}, Burger},
        title = "{Feature guided training and rotational standardization for the morphological classification of radio galaxies}",
      journal = {\mnras},
         year = 2023,
        
       volume = {522},
       number = {1},
        pages = {292-311},
          doi = {https://doi.org/10.1093/mnras/stad989},
}

@article{polsterer2019pink,
  title={Pink: Parallelized rotation and flipping invariant kohonen maps},
  author={Polsterer, Kai Lars and Gieseke, Fabian and Doser, Bernd},
  journal={Astrophysics Source Code Library},
  pages={ascl--1910},
  year={2019}
}

@ARTICLE{Lukic18,
       author = {{Lukic}, V. and {Br{\"u}ggen}, M. and {Banfield}, J.~K. and {Wong}, O.~I. and {Rudnick}, L. and {Norris}, R.~P. and {Simmons}, B.},
        title = "{Radio Galaxy Zoo: compact and extended radio source classification with deep learning}",
      journal = {\mnras},
         year = 2018,
        
       volume = {476},
       number = {1},
        pages = {246-260},
          doi = {https://doi.org/10.1093/mnras/sty163}
}

@ARTICLE{Alhassan18,
       author = {{Alhassan}, Wathela and {Taylor}, A.~R. and {Vaccari}, Mattia},
        title = "{The FIRST Classifier: compact and extended radio galaxy classification using deep Convolutional Neural Networks}",
      journal = {\mnras},
         year = 2018,
        
       volume = {480},
       number = {2},
        pages = {2085-2093},
          doi = {https://doi.org/10.1093/mnras/sty2038}
}

@ARTICLE{Aniyan17,
       author = {{Aniyan}, A.~K. and {Thorat}, K.},
        title = "{Classifying Radio Galaxies with the Convolutional Neural Network}",
      journal = {\apjs},
         year = 2017,
        
       volume = {230},
       number = {2},
          eid = {20},
        pages = {20},
          doi = {https://doi.org/10.3847/1538-4365/aa7333}
}

@ARTICLE{Becker21,
       author = {{Becker}, Burger and {Vaccari}, Mattia and {Prescott}, Matthew and {Grobler}, Trienko},
        title = "{CNN architecture comparison for radio galaxy classification}",
      journal = {\mnras},
         year = 2021,
        
       volume = {503},
       number = {2},
        pages = {1828-1846},
          doi = {https://doi.org/10.1093/mnras/stab325}
}

@ARTICLE{Tang22,
       author = {{Tang}, H. and {Scaife}, A.~M.~M. and {Wong}, O.~I. and {Shabala}, S.~S.},
        title = "{Radio Galaxy Zoo: giant radio galaxy classification using multidomain deep learning}",
      journal = {\mnras},
         year = 2022,
        
       volume = {510},
       number = {3},
        pages = {4504-4524},
          doi = {https://doi.org/10.1093/mnras/stab3553}
}

@ARTICLE{Scaife21,
       author = {{Scaife}, Anna M.~M. and {Porter}, Fiona},
        title = "{Fanaroff-Riley classification of radio galaxies using group-equivariant convolutional neural networks}",
      journal = {\mnras},
         year = 2021,
        
       volume = {503},
       number = {2},
        pages = {2369-2379},
          doi = {https://doi.org/10.1093/mnras/stab530}
}

@inproceedings{cohen16,
  title={Group equivariant convolutional networks},
  author={Cohen, Taco and Welling, Max},
  booktitle={International conference on machine learning},
  pages={2990-2999},
  year={2016},
  organization={PMLR}
}

@article{krizhevsky2012imagenet,
  title={Imagenet classification with deep convolutional neural networks},
  author={Krizhevsky, Alex and Sutskever, Ilya and Hinton, Geoffrey E},
  journal={Advances in neural information processing systems},
  volume={25},
  year={2012}
}

@ARTICLE{Bowles21,
       author = {{Bowles}, Micah and {Scaife}, Anna M.~M. and {Porter}, Fiona and {Tang}, Hongming and {Bastien}, David J.},
        title = "{Attention-gating for improved radio galaxy classification}",
      journal = {\mnras},
         year = 2021,
        
       volume = {501},
       number = {3},
        pages = {4579-4595},
          doi = {https://doi.org/10.1093/mnras/staa3946}
}

@ARTICLE{Rustige23,
       author = {{Rustige}, Lennart and {Kummer}, Janis and {Griese}, Florian and {Borras}, Kerstin and {Br{\"u}ggen}, Marcus and {Connor}, Patrick L.~S. and {Gaede}, Frank and {Kasieczka}, Gregor and {Knopp}, Tobias and {Schleper}, Peter},
        title = "{Morphological classification of radio galaxies with Wasserstein generative adversarial network-supported augmentation}",
      journal = {RAS Techniques and Instruments},
         year = 2023,
        
       volume = {2},
       number = {1},
        pages = {264-277},
          doi = {https://doi.org/10.1093/rasti/rzad016}
}

@ARTICLE{alger18,
       author = {{Alger}, M.~J. and {Banfield}, J.~K. and {Ong}, C.~S. and {Rudnick}, L. and {Wong}, O.~I. and {Wolf}, C. and {Andernach}, H. and {Norris}, R.~P. and {Shabala}, S.~S.},
        title = "{Radio Galaxy Zoo: machine learning for radio source host galaxy cross-identification}",
      journal = {\mnras},
         year = 2018,
        
       volume = {478},
       number = {4},
        pages = {5547},
          doi = {https://doi.org/10.1093/mnras/sty1308},
archivePrefix = {arXiv},
       eprint = {1805.05540},
 primaryClass = {astro-ph.IM},
       adsurl = {https://ui.adsabs.harvard.edu/abs/2018MNRAS.478.5547A}
}

@ARTICLE{lao21,
       author = {{Lao}, Baoqiang and {An}, Tao and {Wang}, Ailing and {Xu}, Zhijun and {Guo}, Shaoguang and {Lv}, Weijia and {Wu}, Xiaocong and {Zhang}, Yingkang},
        title = "{Artificial intelligence for celestial object census: the latest technology meets the oldest science}",
      journal = {Science Bulletin},
         year = 2021,
        
       volume = {66},
       number = {21},
        pages = {2145},
          doi = {https://doi.org/10.1016/j.scib.2021.07.015},
archivePrefix = {arXiv},
       eprint = {2107.03082},
 primaryClass = {astro-ph.IM},
       adsurl = {https://ui.adsabs.harvard.edu/abs/2021SciBu..66.2145L}
}

@misc{magro_2022,
  AUTHOR = {Magro, D. and Riggi, S. and {De Marco}, A. and Adami, K. Z. and Sortino, R. and Pino, C. and Bordiu, C. and Cecconello, T. and Sciacca, E. and Umana, G. and Ingallinra, A. and Bufano, F. and Spampinato, C. and Vizzari, G. and Vitello, F. and Becciani, U. and Collier, J. D. and Marvil, J. and Michalowski, M. J. and Hopkins, A. M.},
  TITLE = {ASGARD: A deep neural network for the detection of compact sources, extended galaxies, and sidelobes in radio astronomical maps},
  NOTE  = {In preparation},
  YEAR  = {2022}
}

@article{huynh_2012,
  title={The Completeness and Reliability of Threshold and False-discovery Rate Source Extraction Algorithms for Compact Continuum Sources},
  author={{Huynh}, M. T. and {Hopkins}, A. and {Norris}, R. and {Hancock}, P. and {Murphy}, T. and {Jurek}, R. and {Whitting}, M.},
  journal={\pasa},
  volume={29},
  pages={229},
  year={2012},
  doi={https://doi.org/10.1071/AS11026}
}

@article{kron_1980,
  title={Photometry of a complete sample of faint galaxies},
  author={{Kron}, R.~G.},
  journal={\apjs},
  volume={43},
  year={1980},
  pages={305},
  doi={https://doi.org/10.1086/190669}
}

@article{samudre2022data,
  title={Data-efficient classification of radio galaxies},
  author={Samudre, Ashwin and George, Lijo T and Bansal, Mahak and Wadadekar, Yogesh},
  journal={Monthly Notices of the Royal Astronomical Society},
  volume={509},
  number={2},
  pages={2269-2280},
  year={2022},
  publisher={Oxford University Press},
  doi = {https://doi.org/10.1093/mnras/stz1883}
}

@ARTICLE{tang2019,
       author = {{Tang}, H. and {Scaife}, A.~M.~M. and {Leahy}, J.~P.},
        title = "{Transfer learning for radio galaxy classification}",
      journal = {\mnras},
         year = 2019,
        
       volume = {488},
       number = {3},
        pages = {3358-3375},
          doi = {https://doi.org/10.1093/mnras/stz1883}
}

@inproceedings{szegedy2017inception,
  title={Inception-v4, inception-resnet and the impact of residual connections on learning},
  author={Szegedy, Christian and Ioffe, Sergey and Vanhoucke, Vincent and Alemi, Alexander},
  booktitle={Proceedings of the AAAI conference on artificial intelligence},
  volume={31},
  number={1},
  year={2017}
}

@inproceedings{darya2023morphological,
  title={Morphological classification of extragalactic radio sources using gradient boosting methods},
  author={Darya, Abdollah Masoud and Fernini, Ilias and Vellasco, Marley and Hussain, Abir},
  booktitle={2023 International Joint Conference on Neural Networks (IJCNN)},
  pages={1-8},
  year={2023},
  organization={IEEE}
}

@ARTICLE{Lukic2019b,
       author = {{Lukic}, V. and {Br{\"u}ggen}, M. and {Mingo}, B. and {Croston}, J.~H. and {Kasieczka}, G. and {Best}, P.~N.},
        title = "{Morphological classification of radio galaxies: capsule networks versus convolutional neural networks}",
      journal = {\mnras},
         year = 2019,
        
       volume = {487},
       number = {2},
        pages = {1729-1744},
          doi = {https://doi.org/10.1093/mnras/stz1289}
}

@article{petrosian_1976,
  title={Surface Brightness and Evolution of Galaxies},
  author={Petrosian, V.},
  journal={Astrophys. J. Lett.},
  year={1976},
  volume={210},
  pages={L53},
  doi={https://doi.org/10.1086/182301}
}

@ARTICLE{Barkai2023,
       author = {{Barkai}, J.~A. and {Verheijen}, M.~A.~W. and {Talavera}, E. and {Wilkinson}, M.~H.~F.},
        title = "{A comparative study of source-finding techniques in H I emission line cubes using SoFiA, MTObjects, and supervised deep learning}",
      journal = {\aap},
         year = 2023,
        
       volume = {670},
          eid = {A55},
        pages = {A55},
          doi = {https://doi.org/10.1051/0004-6361/202244708},
archivePrefix = {arXiv},
       eprint = {2211.12809},
 primaryClass = {astro-ph.IM},
       adsurl = {https://ui.adsabs.harvard.edu/abs/2023A&A...670A..55B}
}

@ARTICLE{hartley23,
       author = {{Hartley}, P. and {Bonaldi}, A. and {Braun}, R. and {Aditya}, J.~N.~H.~S. and {Aicardi}, S. and {Alegre}, L. and {Chakraborty}, A. and {Chen}, X. and {Choudhuri}, S. and {Clarke}, A.~O. and {Coles}, J. and {Collinson}, J.~S. and {Cornu}, D. and {Darriba}, L. and {Veneri}, M. Delli and {Forbrich}, J. and {Fraga}, B. and {Galan}, A. and {Garrido}, J. and {Gubanov}, F. and {H{\r{a}}kansson}, H. and {Hardcastle}, M.~J. and {Heneka}, C. and {Herranz}, D. and {Hess}, K.~M. and {Jagannath}, M. and {Jaiswal}, S. and {Jurek}, R.~J. and {Korber}, D. and {Kitaeff}, S. and {Kleiner}, D. and {Lao}, B. and {Lu}, X. and {Mazumder}, A. and {Mold{\'o}n}, J. and {Mondal}, R. and {Ni}, S. and {{\"O}nnheim}, M. and {Parra}, M. and {Patra}, N. and {Peel}, A. and {Salom{\'e}}, P. and {S{\'a}nchez-Exp{\'o}sito}, S. and {Sargent}, M. and {Semelin}, B. and {Serra}, P. and {Shaw}, A.~K. and {Shen}, A.~X. and {Sj{\"o}berg}, A. and {Smith}, L. and {Soroka}, A. and {Stolyarov}, V. and {Tolley}, E. and {Toribio}, M.~C. and {van der Hulst}, J.~M. and {Sadr}, A. Vafaei and {Verdes-Montenegro}, L. and {Westmeier}, T. and {Yu}, K. and {Yu}, L. and {Zhang}, L. and {Zhang}, X. and {Zhang}, Y. and {Alberdi}, A. and {Ashdown}, M. and {Bom}, C.~R. and {Br{\"u}ggen}, M. and {Cannon}, J. and {Chen}, R. and {Combes}, F. and {Conway}, J. and {Courbin}, F. and {Ding}, J. and {Fourestey}, G. and {Freundlich}, J. and {Gao}, L. and {Gheller}, C. and {Guo}, Q. and {Gustavsson}, E. and {Jirstrand}, M. and {Jones}, M.~G. and {J{\'o}zsa}, G. and {Kamphuis}, P. and {Kneib}, J. -P. and {Lindqvist}, M. and {Liu}, B. and {Liu}, Y. and {Mao}, Y. and {Marchal}, A. and {M{\'a}rquez}, I. and {Meshcheryakov}, A. and {Olberg}, M. and {Oozeer}, N. and {Pandey-Pommier}, M. and {Pei}, W. and {Peng}, B. and {Sabater}, J. and {Sorgho}, A. and {Starck}, J.~L. and {Tasse}, C. and {Wang}, A. and {Wang}, Y. and {Xi}, H. and {Yang}, X. and {Zhang}, H. and {Zhang}, J. and {Zhao}, M. and {Zuo}, S.},
        title = "{SKA Science Data Challenge 2: analysis and results}",
      journal = {Monthly Notices of the Royal Astronomical Society},
         year = 2023,
        
       volume = {523},
       number = {2},
        pages = {1967-1993},
          doi = {https://doi.org/10.1093/mnras/stad1375},
archivePrefix = {arXiv},
       eprint = {2303.07943},
 primaryClass = {astro-ph.IM},
       adsurl = {https://ui.adsabs.harvard.edu/abs/2023MNRAS.523.1967H}
}

@article{haakansson2023utilization,
  title={Utilization of convolutional neural networks for H I source finding-Team FORSKA-Sweden approach to SKA Data Challenge 2},
  author={H{\aa}kansson, Henrik and Sj{\"o}berg, Anders and Toribio, Maria Carmen and {\"O}nnheim, Magnus and Olberg, Michael and Gustavsson, Emil and Lindqvist, Michael and Jirstrand, Mats and Conway, John},
  journal={Astronomy \& Astrophysics},
  volume={671},
  pages={A39},
  year={2023},
  doi = {https://doi.org/10.1051/0004-6361/202245139},
  publisher={EDP Sciences}
}

@ARTICLE{gopal_krishna_2000,
       author = {{Gopal-Krishna} and {Wiita}, P.~J.},
        title = "{Extragalactic radio sources with hybrid morphology: implications for the Fanaroff-Riley dichotomy}",
      journal = {\aap},
         year = 2000,
        
       volume = {363},
        pages = {507-516},
          doi = {https://doi.org/10.48550/arXiv.astro-ph/0009441},
archivePrefix = {arXiv},
       eprint = {astro-ph/0009441},
 primaryClass = {astro-ph},
       adsurl = {https://ui.adsabs.harvard.edu/abs/2000A&A...363..507G}
}

@ARTICLE{harwood20,
       author = {{Harwood}, Jeremy J. and {Vernstrom}, Tessa and {Stroe}, Andra},
        title = "{Unveiling the cause of hybrid morphology radio sources (HyMoRS)}",
      journal = {\mnras},
         year = 2020,
        
       volume = {491},
       number = {1},
        pages = {803-822},
          doi = {https://doi.org/10.1093/mnras/stz3069},
archivePrefix = {arXiv},
       eprint = {1910.12857},
 primaryClass = {astro-ph.GA},
       adsurl = {https://ui.adsabs.harvard.edu/abs/2020MNRAS.491..803H}
}

@ARTICLE{kumari22,
       author = {{Kumari}, Shobha and {Pal}, Sabyasachi},
        title = "{Search for hybrid morphology radio galaxies from the FIRST survey at 1400 MHz}",
      journal = {\mnras},
         year = 2022,
        
       volume = {514},
       number = {3},
        pages = {4290-4299},
          doi = {https://doi.org/10.1093/mnras/stac1215},
archivePrefix = {arXiv},
       eprint = {2104.14469},
 primaryClass = {astro-ph.GA},
       adsurl = {https://ui.adsabs.harvard.edu/abs/2022MNRAS.514.4290K}
}

@article{Manik2026,
  title = {Hybrid Morphology Radio Sources from the MeerKAT Absorption Line Survey (MALS): Radio,  Mid-infrared,  and Environmental Characteristics},
  volume = {997},
  ISSN = {1538-4357},
  DOI = {https://doi.org/10.3847/1538-4357/ae1f85},
  number = {2},
  journal = {The Astrophysical Journal},
  publisher = {American Astronomical Society},
  author = {Manik,  Souvik and Kumari,  Shobha and Bhukta,  Netai and Pal,  Sabyasachi and Mondal,  Sushanta K.},
  year = {2026},
  
  pages = {157}
}

@ARTICLE{molinari_2011,
       author = {{Molinari}, S. and {Schisano}, E. and {Faustini}, F. and {Pestalozzi}, M. and {di Giorgio}, A.~M. and {Liu}, S.},
        title = "{Source extraction and photometry for the far-infrared and sub-millimeter continuum in the presence of complex backgrounds}",
      journal = {\aap},
         year = 2011,
        
       volume = {530},
          eid = {A133},
        pages = {A133},
          doi = {https://doi.org/10.1051/0004-6361/201014752},
archivePrefix = {arXiv},
       eprint = {1011.3946},
 primaryClass = {astro-ph.GA},
       adsurl = {https://ui.adsabs.harvard.edu/abs/2011A&A...530A.133M}
}

@ARTICLE{riggi_2024,
       author = {{Riggi}, S. and {Cecconello}, T. and {Palazzo}, S. and {Hopkins}, A.~M. and {Gupta}, N. and {Bordiu}, C. and {Ingallinera}, A. and {Buemi}, C. and {Bufano}, F. and {Cavallaro}, F. and {Filipovi{\'c}}, M.~D. and {Leto}, P. and {Loru}, S. and {Ruggeri}, A.~C. and {Trigilio}, C. and {Umana}, G. and {Vitello}, F.},
        title = "{Self-supervised contrastive learning of radio data for source detection, classification and peculiar object discovery}",
      journal = {\pasa},
         year = 2024,
        
       volume = {41},
          eid = {e085},
        pages = {e085},
          doi = {https://doi.org/10.1017/pasa.2024.84},
archivePrefix = {arXiv},
       eprint = {2404.18462},
 primaryClass = {astro-ph.IM},
       adsurl = {https://ui.adsabs.harvard.edu/abs/2024PASA...41...85R}
}

@ARTICLE{lastufka_2024,
       author = {{Lastufka}, E. and {Bait}, O. and {Taran}, O. and {Drozdova}, M. and {Kinakh}, V. and {Piras}, D. and {Audard}, M. and {Dessauges-Zavadsky}, M. and {Holotyak}, T. and {Schaerer}, D. and {Voloshynovskiy}, S.},
        title = "{Self-supervised learning on MeerKAT wide-field continuum images}",
      journal = {\aap},
         year = 2024,
        
       volume = {690},
          eid = {A310},
        pages = {A310},
          doi = {https://doi.org/10.1051/0004-6361/202449964},
archivePrefix = {arXiv},
       eprint = {2408.06147},
 primaryClass = {astro-ph.IM},
       adsurl = {https://ui.adsabs.harvard.edu/abs/2024A&A...690A.310L}
}

@ARTICLE{gupta_2023,
       author = {{Gupta}, Nikhel and {Hayder}, Zeeshan and {Norris}, Ray P. and {Huynh}, Minh and {Petersson}, Lars and {Wang}, X. Rosalind and {Andernach}, Heinz and {Koribalski}, B{\"a}rbel S. and {Yew}, Miranda and {Crawford}, Evan J.},
        title = "{Deep learning for morphological identification of extended radio galaxies using weak labels}",
      journal = {\pasa},
         year = 2023,
        
       volume = {40},
          eid = {e044},
        pages = {e044},
          doi = {https://doi.org/10.1017/pasa.2023.46},
archivePrefix = {arXiv},
       eprint = {2308.05166},
 primaryClass = {astro-ph.IM},
       adsurl = {https://ui.adsabs.harvard.edu/abs/2023PASA...40...44G}
}

@ARTICLE{slijepcevic_2022,
       author = {{Slijepcevic}, Inigo V. and {Scaife}, Anna M.~M. and {Walmsley}, Mike and {Bowles}, Micah and {Wong}, O. Ivy and {Shabala}, Stanislav S. and {Tang}, Hongming},
        title = "{Radio Galaxy Zoo: using semi-supervised learning to leverage large unlabelled data sets for radio galaxy classification under data set shift}",
      journal = {\mnras},
         year = 2022,
        
       volume = {514},
       number = {2},
        pages = {2599-2613},
          doi = {https://doi.org/10.1093/mnras/stac1135},
archivePrefix = {arXiv},
       eprint = {2204.08816},
 primaryClass = {astro-ph.GA},
       adsurl = {https://ui.adsabs.harvard.edu/abs/2022MNRAS.514.2599S}
}

@ARTICLE{slijepcevic_2024,
       author = {{Slijepcevic}, Inigo V. and {Scaife}, Anna M.~M. and {Walmsley}, Mike and {Bowles}, Micah and {Wong}, O. Ivy and {Shabala}, Stanislav S. and {White}, Sarah V.},
        title = "{Radio galaxy zoo: towards building the first multipurpose foundation model for radio astronomy with self-supervised learning}",
      journal = {RAS Techniques and Instruments},
         year = 2024,
        
       volume = {3},
       number = {1},
        pages = {19-32},
          doi = {https://doi.org/10.1093/rasti/rzad055},
archivePrefix = {arXiv},
       eprint = {2305.16127},
 primaryClass = {astro-ph.IM},
       adsurl = {https://ui.adsabs.harvard.edu/abs/2024RASTI...3...19S}
}

@ARTICLE{ma_2019,
       author = {{Ma}, Zhixian and {Xu}, Haiguang and {Zhu}, Jie and {Hu}, Dan and {Li}, Weitian and {Shan}, Chenxi and {Zhu}, Zhenghao and {Gu}, Liyi and {Li}, Jinjin and {Liu}, Chengze and {Wu}, Xiangping},
        title = "{A Machine Learning Based Morphological Classification of 14,245 Radio AGNs Selected from the Best-Heckman Sample}",
      journal = {\apjs},
         year = 2019,
        
       volume = {240},
       number = {2},
          eid = {34},
        pages = {34},
          doi = {https://doi.org/10.3847/1538-4365/aaf9a2},
archivePrefix = {arXiv},
       eprint = {1812.07190},
 primaryClass = {astro-ph.GA},
       adsurl = {https://ui.adsabs.harvard.edu/abs/2019ApJS..240...34M}
}

@ARTICLE{cecconello_2024,
       author = {{Cecconello}, Thomas and {Riggi}, Simone and {Becciani}, Ugo and {Vitello}, Fabio and {Hopkins}, Andrew M. and {Vizzari}, Giuseppe and {Spampinato}, Concetto and {Palazzo}, Simone},
        title = "{Self-supervised learning for radio-astronomy source classification: a benchmark}",
      journal = {arXiv e-prints},
         year = 2024,
        
          eid = {arXiv:2411.14078},
        pages = {arXiv:2411.14078},
          doi = {https://doi.org/10.48550/arXiv.2411.14078},
archivePrefix = {arXiv},
       eprint = {2411.14078},
 primaryClass = {astro-ph.IM},
       adsurl = {https://ui.adsabs.harvard.edu/abs/2024arXiv241114078C}
}

@ARTICLE{hossain_2023,
       author = {{Hossain}, Mir Sazzat and {Roy}, Sugandha and {Asad}, K.~M.~B. and {Momen}, Arshad and {Ali}, Amin Ahsan and {Amin}, M. Ashraful and {Rahman}, A.~K.~M. Mahbubur},
        title = "{Morphological Classification of Radio Galaxies using Semi-Supervised Group Equivariant CNNs}",
      journal = {Procedia Computer Science},
         year = 2023,
        
       volume = {222},
        pages = {601-612},
          doi = {https://doi.org/10.1016/j.procs.2023.08.198},
archivePrefix = {arXiv},
       eprint = {2306.00031},
 primaryClass = {astro-ph.IM},
       adsurl = {https://ui.adsabs.harvard.edu/abs/2023PrCS..222..601H}
}

@ARTICLE{fanaroff_riley_1974,
       author = {{Fanaroff}, B.~L. and {Riley}, J.~M.},
        title = "{The morphology of extragalactic radio sources of high and low luminosity}",
      journal = {\mnras},
         year = 1974,
        
       volume = {167},
        pages = {31P-36P},
          doi = {https://doi.org/10.1093/mnras/167.1.31P},
       adsurl = {https://ui.adsabs.harvard.edu/abs/1974MNRAS.167P..31F}
}

@software{jax2018github,
	author = {James Bradbury and Roy Frostig and Peter Hawkins and Matthew James Johnson and Chris Leary and Dougal Maclaurin and George Necula and Adam Paszke and Jake Vander{P}las and Skye Wanderman-{M}ilne and Qiao Zhang},
	title = {{JAX}: composable transformations of {P}ython+{N}um{P}y programs},
	url = {https://github.com/google/jax},
	version = {0.3.13},
	year = {2018},
}

@ARTICLE{Conselice_2000,
	author = {Conselice, C.~J. and Bershady, M.~A. and Jangren, A.},
	title = "{The Asymmetry of Galaxies: Physical Morphology for Nearby and High-Redshift Galaxies}",
	journal = {\apj},
	eprint = {astro-ph/9907399},
	year = 2000,
	
	volume = 529,
	pages = {886-910},
	doi = {https://doi.org/10.1086/308300},
	adsurl = {https://adsabs.harvard.edu/abs/2000ApJ...529..886C}
}

@ARTICLE{Lotz_2004,
	author = {{Lotz}, Jennifer M. and {Primack}, Joel and {Madau}, Piero},
	title = "{A New Nonparametric Approach to Galaxy Morphological Classification}",
	journal = {\aj},
	year = 2004,
	
	volume = {128},
	number = {1},
	pages = {163-182},
	doi = {https://doi.org/10.1086/421849},
	archivePrefix = {arXiv},
	eprint = {astro-ph/0311352},
	primaryClass = {astro-ph},
	adsurl = {https://ui.adsabs.harvard.edu/abs/2004AJ....128..163L}
}

@ARTICLE{Lotz_2008,
	author = {{Lotz}, Jennifer M. and {Davis}, M. and {Faber}, S.~M. and {Guhathakurta}, P. and {Gwyn}, S. and {Huang}, J. and {Koo}, D.~C. and {Le Floc'h}, E. and {Lin}, Lihwai and {Newman}, J. and {Noeske}, K. and {Papovich}, C. and {Willmer}, C.~N.~A. and {Coil}, A. and {Conselice}, C.~J. and {Cooper}, M. and {Hopkins}, A.~M. and {Metevier}, A. and {Primack}, J. and {Rieke}, G. and {Weiner}, B.~J.},
	title = "{The Evolution of Galaxy Mergers and Morphology at z < 1.2 in the Extended Groth Strip}",
	journal = {\apj},
	year = 2008,
	
	volume = {672},
	number = {1},
	pages = {177-197},
	doi = {https://doi.org/10.1086/523659},
	archivePrefix = {arXiv},
	eprint = {astro-ph/0602088},
	primaryClass = {astro-ph},
	adsurl = {https://ui.adsabs.harvard.edu/abs/2008ApJ...672..177L}
}

@ARTICLE{Simard_2002,
	author = {{Simard}, Luc and {Willmer}, Christopher N.~A. and {Vogt}, Nicole P. and {Sarajedini}, Vicki L. and {Phillips}, Andrew C. and {Weiner}, Benjamin J. and {Koo}, David C. and {Im}, Myungshin and {Illingworth}, Garth D. and {Faber}, S.~M.},
	title = "{The DEEP Groth Strip Survey. II. Hubble Space Telescope Structural Parameters of Galaxies in the Groth Strip}",
	journal = {\apjs},
	year = 2002,
	
	volume = {142},
	number = {1},
	pages = {1-33},
	doi = {https://doi.org/10.1086/341399},
	archivePrefix = {arXiv},
	eprint = {astro-ph/0205025},
	primaryClass = {astro-ph},
	adsurl = {https://ui.adsabs.harvard.edu/abs/2002ApJS..142....1S}
}

@ARTICLE{Peng_2010,
	author = {{Peng}, Ying-jie and {Lilly}, Simon J. and {Kova{\v{c}}}, Katarina and {Bolzonella}, Micol and {Pozzetti}, Lucia and {Renzini}, Alvio and {Zamorani}, Gianni and {Ilbert}, Olivier and {Knobel}, Christian and {Iovino}, Angela and {Maier}, Christian and {Cucciati}, Olga and {Tasca}, Lidia and {Carollo}, C. Marcella and {Silverman}, John and {Kampczyk}, Pawel and {de Ravel}, Loic and {Sanders}, David and {Scoville}, Nicholas and {Contini}, Thierry and {Mainieri}, Vincenzo and {Scodeggio}, Marco and {Kneib}, Jean-Paul and {Le F{\`e}vre}, Olivier and {Bardelli}, Sandro and {Bongiorno}, Angela and {Caputi}, Karina and {Coppa}, Graziano and {de la Torre}, Sylvain and {Franzetti}, Paolo and {Garilli}, Bianca and {Lamareille}, Fabrice and {Le Borgne}, Jean-Francois and {Le Brun}, Vincent and {Mignoli}, Marco and {Perez Montero}, Enrique and {Pello}, Roser and {Ricciardelli}, Elena and {Tanaka}, Masayuki and {Tresse}, Laurence and {Vergani}, Daniela and {Welikala}, Niraj and {Zucca}, Elena and {Oesch}, Pascal and {Abbas}, Ummi and {Barnes}, Luke and {Bordoloi}, Rongmon and {Bottini}, Dario and {Cappi}, Alberto and {Cassata}, Paolo and {Cimatti}, Andrea and {Fumana}, Marco and {Hasinger}, Gunther and {Koekemoer}, Anton and {Leauthaud}, Alexei and {Maccagni}, Dario and {Marinoni}, Christian and {McCracken}, Henry and {Memeo}, Pierdomenico and {Meneux}, Baptiste and {Nair}, Preethi and {Porciani}, Cristiano and {Presotto}, Valentina and {Scaramella}, Roberto},
	title = "{Mass and Environment as Drivers of Galaxy Evolution in SDSS and zCOSMOS and the Origin of the Schechter Function}",
	journal = {\apj},
	year = 2010,
	
	volume = {721},
	number = {1},
	pages = {193-221},
	doi = {https://doi.org/10.1088/0004-637X/721/1/193},
	archivePrefix = {arXiv},
	eprint = {1003.4747},
	primaryClass = {astro-ph.CO},
	adsurl = {https://ui.adsabs.harvard.edu/abs/2010ApJ...721..193P}
}

@article{loreto2015,
	author = {{Barcos-Mu{\~n}oz}, L. and {Leroy}, A.~K. and {Evans}, A.~S. and {Privon}, G.~C. and {Armus}, L. and {Condon}, J. and {Mazzarella}, J.~M. and {Meier}, D.~S. and {Momjian}, E. and {Murphy}, E.~J. and {Ott}, J. and {Reichardt}, A. and {Sakamoto}, K. and {Sanders}, D.~B. and {Schinnerer}, E. and {Stierwalt}, S. and {Surace}, J.~A. and {Thompson}, T.~A. and {Walter}, F.},
	title = "{High-resolution Radio Continuum Measurements of the Nuclear Disks of Arp 220}",
	journal = {\apj},
	year = 2015,
	
	volume = {799},
	number = {1},
	eid = {10},
	pages = {10},
	doi = {https://doi.org/10.1088/0004-637X/799/1/10},
	archivePrefix = {arXiv},
	eprint = {1411.0932},
	primaryClass = {astro-ph.GA},
	adsurl = {https://ui.adsabs.harvard.edu/abs/2015ApJ...799...10B}
}

@ARTICLE{Hodge2019,
	author = {{Hodge}, J.~A. and {Smail}, I. and {Walter}, F. and {da Cunha}, E. and {Swinbank}, A.~M. and {Rybak}, M. and {Venemans}, B. and {Brandt}, W.~N. and {Calistro Rivera}, G. and {Chapman}, S.~C. and {Chen}, Chian-Chou and {Cox}, P. and {Dannerbauer}, H. and {Decarli}, R. and {Greve}, T.~R. and {Knudsen}, K.~K. and {Menten}, K.~M. and {Schinnerer}, E. and {Simpson}, J.~M. and {van der Werf}, P. and {Wardlow}, J.~L. and {Weiss}, A.},
	title = "{ALMA Reveals Potential Evidence for Spiral Arms, Bars, and Rings in High-redshift Submillimeter Galaxies}",
	journal = {\apj},
	year = 2019,
	
	volume = {876},
	number = {2},
	eid = {130},
	pages = {130},
	doi = {https://doi.org/10.3847/1538-4357/ab1846},
	archivePrefix = {arXiv},
	eprint = {1810.12307},
	primaryClass = {astro-ph.GA},
	adsurl = {https://ui.adsabs.harvard.edu/abs/2019ApJ...876..130H}
}

@article{Song2021,
	author = {{Song}, Y. and {Linden}, S.~T. and {Evans}, A.~S. and {Barcos-Mu{\~n}oz}, L. and {Privon}, G.~C. and {Yoon}, I. and {Murphy}, E.~J. and {Larson}, K.~L. and {D{\'\i}az-Santos}, T. and {Armus}, L. and {Mazzarella}, Joseph M. and {Howell}, J. and {Inami}, H. and {Torres-Alb{\`a}}, N. and {U}, V. and {Charmandaris}, V. and {McKinney}, J. and {Kunneriath}, D. and {Momjian}, E.},
	title = "{A Comparison between Nuclear Ring Star Formation in LIRGs and in Normal Galaxies with the Very Large Array}",
	journal = {\apj},
	year = 2021,
	
	volume = {916},
	number = {2},
	eid = {73},
	pages = {73},
	doi = {https://doi.org/10.3847/1538-4357/ac05c2},
	archivePrefix = {arXiv},
	eprint = {2107.00412},
	primaryClass = {astro-ph.GA},
	adsurl = {https://ui.adsabs.harvard.edu/abs/2021ApJ...916...73S}
}

@article{Condon_1998,
	author = {{Condon}, J.~J. and {Cotton}, W.~D. and {Greisen}, E.~W. and {Yin}, Q.~F. and {Perley}, R.~A. and {Taylor}, G.~B. and {Broderick}, J.~J.},
	title = "{The NRAO VLA Sky Survey}",
	journal = {\aj},
	year = 1998,
	
	volume = {115},
	number = {5},
	pages = {1693-1716},
	doi = {https://doi.org/10.1086/300337},
	adsurl = {https://ui.adsabs.harvard.edu/abs/1998AJ....115.1693C}
}

@article{Condon_1997,
	author = {Condon, J J},
	doi = {https://doi.org/10.1086/133871},
	journal = {Publications of the Astronomical Society of the Pacific},
	number = {732},
	pages = {166},
	publisher = {The Astronomical Society of the Pacific},
	title = {{ERRORS IN ELLIPTICAL GAUSSIAN FITS}},
	volume = {109},
	year = {1997}
}

@ARTICLE{sersic1963,
	author = {{S{\'e}rsic}, J.~L.},
	title = "{Influence of the atmospheric and instrumental dispersion on the brightness distribution in a galaxy}",
	journal = {Boletin de la Asociacion Argentina de Astronomia La Plata Argentina},
	year = 1963,
	volume = {6},
	pages = {41-43},
	adsurl = {https://ui.adsabs.harvard.edu/abs/1963BAAA....6...41S}
}

@article{caon1993,
	author = {{Caon}, N. and {Capaccioli}, M. and {D'Onofrio}, M.},
	title = "{On the shape of the light profiles of early-type galaxies.}",
	journal = {\mnras},
	year = 1993,
	volume = {265},
	pages = {1013-1021},
	doi = {https://doi.org/10.1093/mnras/265.4.1013},
	archivePrefix = {arXiv},
	eprint = {astro-ph/9309013},
	primaryClass = {astro-ph},
	adsurl = {https://ui.adsabs.harvard.edu/abs/1993MNRAS.265.1013C}
}

@ARTICLE{Haussler_2007,
	author = {{H{\"a}ussler}, Boris and {McIntosh}, Daniel H. and {Barden}, Marco and {Bell}, Eric F. and {Rix}, Hans-Walter and {Borch}, Andrea and {Beckwith}, Steven V.~W. and {Caldwell}, John A.~R. and {Heymans}, Catherine and {Jahnke}, Knud and {Jogee}, Shardha and {Koposov}, Sergey E. and {Meisenheimer}, Klaus and {S{\'a}nchez}, Sebastian F. and {Somerville}, Rachel S. and {Wisotzki}, Lutz and {Wolf}, Christian},
	title = "{GEMS: Galaxy Fitting Catalogs and Testing Parametric Galaxy Fitting Codes: GALFIT and GIM2D}",
	journal = {\apjs},
	year = 2007,
	
	volume = {172},
	number = {2},
	pages = {615-633},
	doi = {https://doi.org/10.1086/518836},
	archivePrefix = {arXiv},
	eprint = {0704.2601},
	primaryClass = {astro-ph},
	adsurl = {https://ui.adsabs.harvard.edu/abs/2007ApJS..172..615H}
}

@ARTICLE{Andrae2011,
	author = {{Andrae}, Ren{\'e} and {Jahnke}, Knud and {Melchior}, Peter},
	title = "{Parametrizing arbitrary galaxy morphologies: potentials and pitfalls}",
	journal = {\mnras},
	year = 2011,
	
	volume = {411},
	number = {1},
	pages = {385-401},
	doi = {https://doi.org/10.1111/j.1365-2966.2010.17690.x},
	archivePrefix = {arXiv},
	eprint = {1009.2508},
	primaryClass = {astro-ph.CO},
	adsurl = {https://ui.adsabs.harvard.edu/abs/2011MNRAS.411..385A}
}

@article{Lucatelli_2024,
	author = {{Lucatelli}, Geferson and {Beswick}, Robert J. and {Mold{\'o}n}, Javier and {P{\'e}rez-Torres}, Miguel A. and {Conway}, John E. and {Alberdi}, Antxon and {Romero-Ca{\~n}izales}, Cristina and {Varenius}, Eskil and {Kl{\"o}ckner}, Hans-Rainer and {Barcos-Mu{\~n}oz}, Loreto and {Bondi}, Marco and {Garrington}, Simon T. and {Aalto}, Susanne and {Baan}, Willem A. and {Pihlstr{\"o}m}, Ylva M.},
	title = "{The PARADIGM project I: a multiscale radio morphological analysis of local U/LIRGS}",
	journal = {\mnras},
	year = 2024,
	volume = {529},
	number = {4},
	pages = {4468-4499},
	doi = {https://doi.org/10.1093/mnras/stae744},
	archivePrefix = {arXiv},
	eprint = {2403.16872},
	primaryClass = {astro-ph.GA},
	adsurl = {https://ui.adsabs.harvard.edu/abs/2024MNRAS.529.4468L}
}

@article{manik2025grq,
  title={Unveiling New Giant Radio Quasars from the TGSS Sky and their Large-scale Environment},
  author={Manik, Souvik and Bhukta, Netai and Pal, Sabyasachi and Mondal, Sushanta. K.},
  journal={\apjs},
  volume={281},
  pages={34},
  year={2025},
  doi = {https://doi.org/10.3847/1538-4365/ae0b52},
  publisher={American Astronomical Society}
}

@article{manik2024grs,
  title={Discovery of giant radio sources from TGSS ADR 1: radio, optical, and infrared properties},
  author={Bhukta, Netai and Manik, Souvik and Pal, Sabyasachi and Mondal, Sushanta. K.},
  journal={ApJS},
  volume={273},
  pages={2},
  year={2024},
  doi = {https://doi.org/10.3847/1538-4365/ad5184},
  publisher={American Astronomical Society}
}

@ARTICLE{Dabhade2020,
       author = {{Dabhade}, P. and {R{\"o}ttgering}, H.~J.~A. and {Bagchi}, J. and {Shimwell}, T.~W. and {Hardcastle}, M.~J. and {Sankhyayan}, S. and {Morganti}, R. and {Jamrozy}, M. and {Shulevski}, A. and {Duncan}, K.~J.},
        title = "{Giant radio galaxies in the LOFAR Two-metre Sky Survey. I. Radio and environmental properties}",
      journal = {\aap},
         year = 2020,
        
       volume = {635},
          eid = {A5},
        pages = {A5},
          doi = {https://doi.org/10.1051/0004-6361/201935589},
archivePrefix = {arXiv},
       eprint = {1904.00409},
 primaryClass = {astro-ph.GA},
       adsurl = {https://ui.adsabs.harvard.edu/abs/2020A&A...635A...5D}
}

@ARTICLE{Shimwell17,
       author = {{Shimwell}, T.~W. and {R{\"o}ttgering}, H.~J.~A. and {Best}, P.~N. and {Williams}, W.~L. and {Dijkema}, T.~J. and {de Gasperin}, F. and {Hardcastle}, M.~J. and {Heald}, G.~H. and {Hoang}, D.~N. and {Horneffer}, A. and {Intema}, H. and {Mahony}, E.~K. and {Mandal}, S. and {Mechev}, A.~P. and {Morabito}, L. and {Oonk}, J.~B.~R. and {Rafferty}, D. and {Retana-Montenegro}, E. and {Sabater}, J. and {Tasse}, C. and {van Weeren}, R.~J. and {Br{\"u}ggen}, M. and {Brunetti}, G. and {Chy{\.z}y}, K.~T. and {Conway}, J.~E. and {Haverkorn}, M. and {Jackson}, N. and {Jarvis}, M.~J. and {McKean}, J.~P. and {Miley}, G.~K. and {Morganti}, R. and {White}, G.~J. and {Wise}, M.~W. and {van Bemmel}, I.~M. and {Beck}, R. and {Brienza}, M. and {Bonafede}, A. and {Calistro Rivera}, G. and {Cassano}, R. and {Clarke}, A.~O. and {Cseh}, D. and {Deller}, A. and {Drabent}, A. and {van Driel}, W. and {Engels}, D. and {Falcke}, H. and {Ferrari}, C. and {Fr{\"o}hlich}, S. and {Garrett}, M.~A. and {Harwood}, J.~J. and {Heesen}, V. and {Hoeft}, M. and {Horellou}, C. and {Israel}, F.~P. and {Kapi{\'n}ska}, A.~D. and {Kunert-Bajraszewska}, M. and {McKay}, D.~J. and {Mohan}, N.~R. and {Orr{\'u}}, E. and {Pizzo}, R.~F. and {Prandoni}, I. and {Schwarz}, D.~J. and {Shulevski}, A. and {Sipior}, M. and {Smith}, D.~J.~B. and {Sridhar}, S.~S. and {Steinmetz}, M. and {Stroe}, A. and {Varenius}, E. and {van der Werf}, P.~P. and {Zensus}, J.~A. and {Zwart}, J.~T.~L.},
        title = "{The LOFAR Two-metre Sky Survey. I. Survey description and preliminary data release}",
      journal = {\aap},
         year = 2017,
       volume = {598},
          eid = {A104},
        pages = {A104},
          doi = {https://doi.org/10.1051/0004-6361/201629313}
}

@ARTICLE{Shimwell22,
       author = {{Shimwell}, T.~W. and {Hardcastle}, M.~J. and {Tasse}, C. and {Best}, P.~N. and {R{\"o}ttgering}, H.~J.~A. and {Williams}, W.~L. and {Botteon}, A. and {Drabent}, A. and {Mechev}, A. and {Shulevski}, A. and {van Weeren}, R.~J. and {Bester}, L. and {Br{\"u}ggen}, M. and {Brunetti}, G. and {Callingham}, J.~R. and {Chy{\.z}y}, K.~T. and {Conway}, J.~E. and {Dijkema}, T.~J. and {Duncan}, K. and {de Gasperin}, F. and {Hale}, C.~L. and {Haverkorn}, M. and {Hugo}, B. and {Jackson}, N. and {Mevius}, M. and {Miley}, G.~K. and {Morabito}, L.~K. and {Morganti}, R. and {Offringa}, A. and {Oonk}, J.~B.~R. and {Rafferty}, D. and {Sabater}, J. and {Smith}, D.~J.~B. and {Schwarz}, D.~J. and {Smirnov}, O. and {O'Sullivan}, S.~P. and {Vedantham}, H. and {White}, G.~J. and {Albert}, J.~G. and {Alegre}, L. and {Asabere}, B. and {Bacon}, D.~J. and {Bonafede}, A. and {Bonnassieux}, E. and {Brienza}, M. and {Bilicki}, M. and {Bonato}, M. and {Calistro Rivera}, G. and {Cassano}, R. and {Cochrane}, R. and {Croston}, J.~H. and {Cuciti}, V. and {Dallacasa}, D. and {Danezi}, A. and {Dettmar}, R.~J. and {Di Gennaro}, G. and {Edler}, H.~W. and {En{\ss}lin}, T.~A. and {Emig}, K.~L. and {Franzen}, T.~M.~O. and {Garc{\'\i}a-Vergara}, C. and {Grange}, Y.~G. and {G{\"u}rkan}, G. and {Hajduk}, M. and {Heald}, G. and {Heesen}, V. and {Hoang}, D.~N. and {Hoeft}, M. and {Horellou}, C. and {Iacobelli}, M. and {Jamrozy}, M. and {Jeli{\'c}}, V. and {Kondapally}, R. and {Kukreti}, P. and {Kunert-Bajraszewska}, M. and {Magliocchetti}, M. and {Mahatma}, V. and {Ma{\l}ek}, K. and {Mandal}, S. and {Massaro}, F. and {Meyer-Zhao}, Z. and {Mingo}, B. and {Mostert}, R.~I.~J. and {Nair}, D.~G. and {Nakoneczny}, S.~J. and {Nikiel-Wroczy{\'n}ski}, B. and {Orr{\'u}}, E. and {Pajdosz-{\'S}mierciak}, U. and {Pasini}, T. and {Prandoni}, I. and {van Piggelen}, H.~E. and {Rajpurohit}, K. and {Retana-Montenegro}, E. and {Riseley}, C.~J. and {Rowlinson}, A. and {Saxena}, A. and {Schrijvers}, C. and {Sweijen}, F. and {Siewert}, T.~M. and {Timmerman}, R. and {Vaccari}, M. and {Vink}, J. and {West}, J.~L. and {Wo{\l}owska}, A. and {Zhang}, X. and {Zheng}, J.},
        title = "{The LOFAR Two-metre Sky Survey. V. Second data release}",
      journal = {\aap},
         year = 2022,
        
       volume = {659},
          eid = {A1},
        pages = {A1},
          doi = {https://doi.org/10.1051/0004-6361/202142484}
}

@ARTICLE{bhukta22,
       author = {{Bhukta}, Netai and {Pal}, Sabyasachi and {Mondal}, Sushanta K.},
        title = "{Search for X/Z-shaped radio sources from TGSS ADR 1}",
      journal = {\mnras},
         year = 2022,     
       volume = {512},
       number = {3},
        pages = {4308-4323},
          doi = {https://doi.org/10.1093/mnras/stac447},
}

@ARTICLE{Riggi19,
       author = {{Riggi}, S. and {Vitello}, F. and {Becciani}, U. and {Buemi}, C. and {Bufano}, F. and {Calanducci}, A. and {Cavallaro}, F. and {Costa}, A. and {Ingallinera}, A. and {Leto}, P. and {Loru}, S. and {Norris}, R.~P. and {Schillir{\`o}}, F. and {Sciacca}, E. and {Trigilio}, C. and {Umana}, G.},
        title = "{Cuc(aesar) source finder: Recent developments and testing}",
      journal = {\pasa},
         year = 2019,        
       volume = {36},
          eid = {e037},
        pages = {e037},
          doi = {https://doi.org/10.1017/pasa.2019.29},
archivePrefix = {arXiv},
       eprint = {1909.06116},
 primaryClass = {astro-ph.IM},
       adsurl = {https://ui.adsabs.harvard.edu/abs/2019PASA...36...37R}
}

@ARTICLE{ngupta25,
       author = {{Gupta}, N. and {Kerp}, J. and {Balashev}, S.~A. and {Morelli}, A.~P.~M. and {Combes}, F. and {Krogager}, J.-K. and {Momjian}, E. and {Borgaonkar}, D. and {Deka}, P.~P. and {Emig}, K.~L. and {Jose}, J. and {J{\'o}zsa}, G.~I.~G. and {Kl{\"o}ckner}, H.-R. and {Moodley}, K. and {Muller}, S. and {Noterdaeme}, P. and {Petitjean}, P. and {Wagenveld}, J.~D.},
        title = "{The MeerKAT Absorption Line Survey (MALS) data release 3: Cold atomic gas associated with the Milky Way}",
      journal = {\aap},
         year = 2025,
        
       volume = {698},
          eid = {A120},
        pages = {A120},
          doi = {https://doi.org/10.1051/0004-6361/202452407},
archivePrefix = {arXiv},
       eprint = {2504.00097},
 primaryClass = {astro-ph.GA},
       adsurl = {https://ui.adsabs.harvard.edu/abs/2025A&A...698A.120G}
}

@ARTICLE{Allison12,
       author = {{Allison}, J.~R. and {Sadler}, E.~M. and {Whiting}, M.~T.},
        title = "{Application of a Bayesian Method to Absorption Spectral-Line Finding in Simulated ASKAP Data}",
      journal = {\pasa},
         year = 2012,
        
       volume = {29},
       number = {3},
        pages = {221-228},
          doi = {https://doi.org/10.1071/AS11040}
}

@ARTICLE{Huang16,
       author = {{Huang}, Gao and {Liu}, Zhuang and {van der Maaten}, Laurens and {Weinberger}, Kilian Q.},
        title = "{Densely Connected Convolutional Networks}",
      journal = {arXiv e-prints},
         year = 2016,
        
          eid = {arXiv:1608.06993},
        pages = {arXiv:1608.06993},
          doi = {https://doi.org/10.48550/arXiv.1608.06993},
archivePrefix = {arXiv},
       eprint = {1608.06993},
 primaryClass = {cs.CV},
       adsurl = {https://ui.adsabs.harvard.edu/abs/2016arXiv160806993H}
}

@ARTICLE{Bera20,
       author = {{Bera}, Soumen and {Pal}, Sabyasachi and {Sasmal}, Tapan K. and {Mondal}, Soumen},
        title = "{FIRST Winged Radio Galaxies with X and Z Symmetry}",
      journal = {\apjs},
         year = 2020,
        
       volume = {251},
       number = {1},
          eid = {9},
        pages = {9},
          doi = {https://doi.org/10.3847/1538-4365/abb367},
archivePrefix = {arXiv},
       eprint = {2011.03839},
 primaryClass = {astro-ph.GA},
       adsurl = {https://ui.adsabs.harvard.edu/abs/2020ApJS..251....9B}
}

@ARTICLE{Bh22tailed,
       author = {{Bhukta}, Netai and {Mondal}, Sushanta K. and {Pal}, Sabyasachi},
        title = "{Tailed radio galaxies from the TIFR GMRT sky survey}",
      journal = {\mnras},
         year = 2022,
        
       volume = {516},
       number = {1},
        pages = {372-390},
          doi = {https://doi.org/10.1093/mnras/stac2001},
archivePrefix = {arXiv},
       eprint = {2110.05484},
 primaryClass = {astro-ph.GA},
       adsurl = {https://ui.adsabs.harvard.edu/abs/2022MNRAS.516..372B}
}

@ARTICLE{Missaglia2019,
       author = {{Missaglia}, V. and {Massaro}, F. and {Capetti}, A. and {Paolillo}, M. and {Kraft}, R.~P. and {Baldi}, R.~D. and {Paggi}, A.},
        title = "{WATCAT: a tale of wide-angle tailed radio galaxies}",
      journal = {\aap},
         year = 2019,
        
       volume = {626},
          eid = {A8},
        pages = {A8},
          doi = {https://doi.org/10.1051/0004-6361/201935058},
archivePrefix = {arXiv},
       eprint = {1904.02175},
 primaryClass = {astro-ph.HE},
       adsurl = {https://ui.adsabs.harvard.edu/abs/2019A&A...626A...8M}
}

@ARTICLE{Konar2013,
       author = {{Konar}, C. and {Hardcastle}, M.~J.},
        title = "{Particle acceleration and dynamics of double-double radio galaxies: theory versus observations}",
      journal = {\mnras},
         year = 2013,
        
       volume = {436},
       number = {2},
        pages = {1595-1614},
          doi = {https://doi.org/10.1093/mnras/stt1676},
archivePrefix = {arXiv},
       eprint = {1309.1401},
 primaryClass = {astro-ph.CO},
       adsurl = {https://ui.adsabs.harvard.edu/abs/2013MNRAS.436.1595K}
}

@ARTICLE{norris2021,
       author = {{Norris}, Ray P. and {Intema}, Huib T. and {Kapi{\'n}ska}, Anna D. and {Koribalski}, B{\"a}rbel S. and {Lenc}, Emil and {Rudnick}, L. and {Alsaberi}, Rami Z.~E. and {Anderson}, Craig and {Anderson}, G.~E. and {Crawford}, E. and {Crocker}, Roland and {English}, Jayanne and {Filipovi{\'c}}, Miroslav D. and {Galvin}, Tim J. and {Hopkins}, Andrew M. and {Hurley-Walker}, Natasha and {Inoue}, Susumu and {Luken}, Kieran and {Macgregor}, Peter J. and {Manojlovi{\'c}}, Pero and {Marvil}, Josh and {O'Brien}, Andrew N. and {Park}, Laurence and {Raja}, Wasim and {Shobhana}, Devika and {Venturi}, Tiziana and {Collier}, Jordan D. and {Hale}, Catherine and {Hotan}, Aidan and {Moss}, Vanessa and {Whiting}, Matthew},
        title = "{Unexpected circular radio objects at high Galactic latitude}",
      journal = {\pasa},
         year = 2021,
        
       volume = {38},
          eid = {e003},
        pages = {e003},
          doi = {https://doi.org/10.1017/pasa.2020.52},
archivePrefix = {arXiv},
       eprint = {2006.14805},
 primaryClass = {astro-ph.GA},
       adsurl = {https://ui.adsabs.harvard.edu/abs/2021PASA...38....3N}
}

@article{manik2025by,
       author = {{Kumari}, Shobha and {Pal}, Sabyasachi and {Manik}, Souvik},
        title = "{Circular Mystery: Exploring Diffuse Emission Surrounding a Radio Galaxy with uGMRT and VLA Multiwavelength Observations}",
      journal = {\apj},
         year = 2025,
        
       volume = {987},
       number = {1},
          eid = {10},
        pages = {10},
          doi = {https://doi.org/10.3847/1538-4357/add69d}
}

@INPROCEEDINGS{2020ASPC..527..635J,
       author = {{J{\'o}zsa}, G.~I.~G. and {White}, S.~V. and {Thorat}, K. and {Smirnov}, O.~M. and {Serra}, P. and {Ramatsoku}, M. and {Ramaila}, A.~J.~T. and {Perkins}, S.~J. and {Maccagni}, F.~M. and {Makhathini}, S. and {Moln{\'a}r}, D.~C. and {Kamphuis}, P. and {Kleiner}, D. and {Hugo}, B.~V. and {de Blok}, W.~J.~G. and {Andati}, L.~A.~L.},
        title = "{MeerKATHI - an End-to-End Data Reduction Pipeline for MeerKAT and Other Radio Telescopes}",
    booktitle = {Astronomical Data Analysis Software and Systems XXIX},
         year = 2020,
       editor = {{Pizzo}, R. and {Deul}, E.~R. and {Mol}, J.~D. and {de Plaa}, J. and {Verkouter}, H.},
       series = {Astronomical Society of the Pacific Conference Series},
       volume = {527},
        
        pages = {635},
          doi = {https://doi.org/10.48550/arXiv.2006.02955},
archivePrefix = {arXiv},
       eprint = {2006.02955},
 primaryClass = {astro-ph.IM},
       adsurl = {https://ui.adsabs.harvard.edu/abs/2020ASPC..527..635J}
}

@INPROCEEDINGS{2024IAUGA..32P1735R,
       author = {{Ramaila}, Athanaseus and {Smirnov}, Oleg and {Hugo}, Benjamin and {Perkins}, Simon and {Ramatsoku}, Mpati and {Andati}, Lexy and {Serra}, Paolo and {Maccagni}, Filippo and {Moln{\'a}r}, D{\'a}niel and {Loi}, Francesca and {White}, Sarah and {Makhathini}, Sphesihle and {Kleiner}, Dane and {de Blok}, Erwin and {Jozsa}, Gyula and {Kamphuis}, Peter and {Thorat}, Kshitij},
        title = "{CARACal: A Comprehensive Container-based Data Reduction Pipeline for Radio Astronomy}",
    booktitle = {32nd General Assembly International Union (IAUGA 2024)},
         year = 2024,
        
          eid = {1735},
        pages = {1735},
       adsurl = {https://ui.adsabs.harvard.edu/abs/2024IAUGA..32P1735R}
}

@INPROCEEDINGS{MIGHTEE,
       author = {{Jarvis}, M. and {Taylor}, R. and {Agudo}, I. and {Allison}, J.~R. and {Deane}, R.~P. and {Frank}, B. and {Gupta}, N. and {Heywood}, I. and {Maddox}, N. and {McAlpine}, K. and {Santos}, M. and {Scaife}, A.~M.~M. and {Vaccari}, M. and {Zwart}, J.~T.~L. and {Adams}, E. and {Bacon}, D.~J. and {Baker}, A.~J. and {Bassett}, B.~A. and {Best}, P.~N. and {Beswick}, R. and {Blyth}, S. and {Brown}, M.~L. and {Bruggen}, M. and {Cluver}, M. and {Colafrancesco}, S. and {Cotter}, G. and {Cress}, C. and {Dav{\'e}}, R. and {Ferrari}, C. and {Hardcastle}, M.~J. and {Hale}, C.~L. and {Harrison}, I. and {Hatfield}, P.~W. and {Klockner}, H.~R. and {Kolwa}, S. and {Malefahlo}, E. and {Marubini}, T. and {Mauch}, T. and {Moodley}, K. and {Morganti}, R. and {Norris}, R.~P. and {Peters}, J.~A. and {Prandoni}, I. and {Prescott}, M. and {Oliver}, S. and {Oozeer}, N. and {Rottgering}, H.~J.~A. and {Seymour}, N. and {Simpson}, C. and {Smirnov}, O. and {Smith}, D.~J.~B.},
        title = "{The MeerKAT International GHz Tiered Extragalactic Exploration (MIGHTEE) Survey}",
    booktitle = {MeerKAT Science: On the Pathway to the SKA},
         year = 2016,        
          eid = {6},
        pages = {6},
          doi = {https://doi.org/10.22323/1.277.0006}
}

@ARTICLE{Hale25,
       author = {{Hale}, C.~L. and {Heywood}, I. and {Jarvis}, M.~J. and {Whittam}, I.~H. and {Best}, P.~N. and {An}, Fangxia and {Bowler}, R.~A.~A. and {Harrison}, I. and {Matthews}, A. and {Smith}, D.~J.~B. and {Taylor}, A.~R. and {Vaccari}, M.},
        title = "{MIGHTEE: the continuum survey Data Release 1}",
      journal = {\mnras},
         year = 2025,
        
       volume = {536},
       number = {3},
        pages = {2187-2211},
          doi = {https://doi.org/10.1093/mnras/stae2528},
archivePrefix = {arXiv},
       eprint = {2411.04958},
 primaryClass = {astro-ph.GA},
       adsurl = {https://ui.adsabs.harvard.edu/abs/2025MNRAS.536.2187H}
}

@ARTICLE{Norris21,
       author = {{Norris}, Ray P. and {Marvil}, Joshua and {Collier}, J.~D. and {Kapi{\'n}ska}, Anna D. and {O'Brien}, Andrew N. and {Rudnick}, L. and {Andernach}, Heinz and {Asorey}, Jacobo and {Brown}, Michael J.~I. and {Br{\"u}ggen}, Marcus and {Crawford}, Evan and {English}, Jayanne and {Rahman}, Syed Faisal ur and {Filipovi{\'c}}, Miroslav D. and {Gordon}, Yjan and {G{\"u}rkan}, G{\"u}lay and {Hale}, Catherine and {Hopkins}, Andrew M. and {Huynh}, Minh T. and {HyeongHan}, Kim and {James Jee}, M. and {Koribalski}, B{\"a}rbel S. and {Lenc}, Emil and {Luken}, Kieran and {Parkinson}, David and {Prandoni}, Isabella and {Raja}, Wasim and {Reiprich}, Thomas H. and {Riseley}, Christopher J. and {Shabala}, Stanislav S. and {Sheil}, Jaimie R. and {Vernstrom}, Tessa and {Whiting}, Matthew T. and {Allison}, James R. and {Anderson}, C.~S. and {Ball}, Lewis and {Bell}, Martin and {Bunton}, John and {Galvin}, T.~J. and {Gupta}, Neeraj and {Hotan}, Aidan and {Jacka}, Colin and {Macgregor}, Peter J. and {Mahony}, Elizabeth K. and {Maio}, Umberto and {Moss}, Vanessa and {Pandey-Pommier}, M. and {Voronkov}, Maxim A.},
        title = "{The Evolutionary Map of the Universe pilot survey}",
      journal = {\pasa},
         year = 2021,
        
       volume = {38},
          eid = {e046},
        pages = {e046},
          doi = {https://doi.org/10.1017/pasa.2021.42},
archivePrefix = {arXiv},
       eprint = {2108.00569},
 primaryClass = {astro-ph.CO},
       adsurl = {https://ui.adsabs.harvard.edu/abs/2021PASA...38...46N}
}

@ARTICLE{Becker95,
       author = {{Becker}, Robert H. and {White}, Richard L. and {Helfand}, David J.},
        title = "{The FIRST Survey: Faint Images of the Radio Sky at Twenty Centimeters}",
      journal = {\apj},
         year = 1995,
        
       volume = {450},
        pages = {559},
          doi = {https://doi.org/10.1086/176166},
       adsurl = {https://ui.adsabs.harvard.edu/abs/1995ApJ...450..559B}
}

@ARTICLE{Gaw06,
       author = {{Gawro{\'n}ski}, M.~P. and {Marecki}, A. and {Kunert-Bajraszewska}, M. and {Kus}, A.~J.},
        title = "{Hybrid morphology radio sources from the FIRST survey}",
      journal = {\aap},
         year = 2006,
        
       volume = {447},
       number = {1},
        pages = {63-70},
          doi = {https://doi.org/10.1051/0004-6361:20053996},
archivePrefix = {arXiv},
       eprint = {astro-ph/0509497},
 primaryClass = {astro-ph},
       adsurl = {https://ui.adsabs.harvard.edu/abs/2006A&A...447...63G}
}

@ARTICLE{kapinska17,
       author = {{Kapi{\'n}ska}, A.~D. and {Terentev}, I. and {Wong}, O.~I. and {Shabala}, S.~S. and {Andernach}, H. and {Rudnick}, L. and {Storer}, L. and {Banfield}, J.~K. and {Willett}, K.~W. and {de Gasperin}, F. and {Lintott}, C.~J. and {L{\'o}pez-S{\'a}nchez}, {\'A}. R. and {Middelberg}, E. and {Norris}, R.~P. and {Schawinski}, K. and {Seymour}, N. and {Simmons}, B.},
        title = "{Radio Galaxy Zoo: A Search for Hybrid Morphology Radio Galaxies}",
      journal = {\aj},
         year = 2017,
        
       volume = {154},
       number = {6},
          eid = {253},
        pages = {253},
          doi = {https://doi.org/10.3847/1538-3881/aa90b7},
archivePrefix = {arXiv},
       eprint = {1711.09611},
 primaryClass = {astro-ph.GA},
       adsurl = {https://ui.adsabs.harvard.edu/abs/2017AJ....154..253K}
}

@ARTICLE{ishwara99,
       author = {{Ishwara-Chandra}, C.~H. and {Saikia}, D.~J.},
        title = "{Giant radio sources}",
      journal = {\mnras},
         year = 1999,
        
       volume = {309},
       number = {1},
        pages = {100-112},
          doi = {https://doi.org/10.1046/j.1365-8711.1999.02835.x},
archivePrefix = {arXiv},
       eprint = {astro-ph/9902252},
 primaryClass = {astro-ph},
       adsurl = {https://ui.adsabs.harvard.edu/abs/1999MNRAS.309..100I}
}

@ARTICLE{oei24,
       author = {{Oei}, Martijn S.~S.~L. and {Hardcastle}, Martin J. and {Timmerman}, Roland and {Gast}, Aivin R.~D.~J.~G.~I.~B. and {Botteon}, Andrea and {Rodriguez}, Antonio C. and {Stern}, Daniel and {Calistro Rivera}, Gabriela and {van Weeren}, Reinout J. and {R{\"o}ttgering}, Huub J.~A. and {Intema}, Huib T. and {de Gasperin}, Francesco and {Djorgovski}, S.~G.},
        title = "{Black hole jets on the scale of the cosmic web}",
      journal = {\nat},
         year = 2024,
        
       volume = {633},
       number = {8030},
        pages = {537-541},
          doi = {https://doi.org/10.1038/s41586-024-07879-y},
archivePrefix = {arXiv},
       eprint = {2411.08630},
 primaryClass = {astro-ph.HE},
       adsurl = {https://ui.adsabs.harvard.edu/abs/2024Natur.633..537O}
}

@ARTICLE{yang19,
       author = {{Yang}, Xiaolong and {Joshi}, Ravi and {Gopal-Krishna} and {An}, Tao and {Ho}, Luis C. and {Wiita}, Paul J. and {Liu}, Xiang and {Yang}, Jun and {Wang}, Ran and {Wu}, Xue-Bing and {Yang}, Xiaofeng},
        title = "{Extended Catalog of Winged or X-shaped Radio Sources from the FIRST Survey}",
      journal = {\apjs},
         year = 2019,
        
       volume = {245},
       number = {1},
          eid = {17},
        pages = {17},
          doi = {https://doi.org/10.3847/1538-4365/ab4811},
archivePrefix = {arXiv},
       eprint = {1905.06356},
 primaryClass = {astro-ph.HE},
       adsurl = {https://ui.adsabs.harvard.edu/abs/2019ApJS..245...17Y}
}

@ARTICLE{Akeret_2017,
       author = {{Akeret}, J. and {Chang}, C. and {Lucchi}, A. and {Refregier}, A.},
        title = "{Radio frequency interference mitigation using deep convolutional neural networks}",
      journal = {Astronomy and Computing},
         year = 2017,
        
       volume = {18},
        pages = {35-39},
          doi = {https://doi.org/10.1016/j.ascom.2017.01.002},
archivePrefix = {arXiv},
       eprint = {1609.09077},
 primaryClass = {astro-ph.IM},
       adsurl = {https://ui.adsabs.harvard.edu/abs/2017A&C....18...35A}
}

@ARTICLE{Mingo_2019,
       author = {{Mingo}, B. and {Croston}, J.~H. and {Hardcastle}, M.~J. and {Best}, P.~N. and {Duncan}, K.~J. and {Morganti}, R. and {Rottgering}, H.~J.~A. and {Sabater}, J. and {Shimwell}, T.~W. and {Williams}, W.~L. and {Brienza}, M. and {Gurkan}, G. and {Mahatma}, V.~H. and {Morabito}, L.~K. and {Prandoni}, I. and {Bondi}, M. and {Ineson}, J. and {Mooney}, S.},
        title = "{Revisiting the Fanaroff-Riley dichotomy and radio-galaxy morphology with the LOFAR Two-Metre Sky Survey (LoTSS)}",
      journal = {\mnras},
         year = 2019,
        
       volume = {488},
       number = {2},
        pages = {2701-2721},
          doi = {https://doi.org/10.1093/mnras/stz1901},
archivePrefix = {arXiv},
       eprint = {1907.03726},
 primaryClass = {astro-ph.GA},
       adsurl = {https://ui.adsabs.harvard.edu/abs/2019MNRAS.488.2701M}
}

@ARTICLE{Rudnick_2021,
       author = {{Rudnick}, Lawrence},
        title = "{Radio Galaxy Classification: \#Tags, Not Boxes}",
      journal = {Galaxies},
         year = 2021,
        
       volume = {9},
       number = {4},
          eid = {85},
        pages = {85},
          doi = {https://doi.org/10.3390/galaxies9040085},
archivePrefix = {arXiv},
       eprint = {2110.13733},
 primaryClass = {astro-ph.GA},
       adsurl = {https://ui.adsabs.harvard.edu/abs/2021Galax...9...85R}
}

@ARTICLE{ContinUNet,
       author = {{Stewart}, Hattie and {Birkinshaw}, Mark and {Yeung}, Siu-Lun and {Maddox}, Natasha and {Maughan}, Ben and {Thiyagalingam}, Jeyan},
        title = "{ContinUNet: fast deep radio image segmentation in the Square Kilometre Array era with U-Net}",
      journal = {RAS Techniques and Instruments},
         year = 2024,
        
       volume = {3},
       number = {1},
        pages = {315-332},
          doi = {https://doi.org/10.1093/rasti/rzae019},
       adsurl = {https://ui.adsabs.harvard.edu/abs/2024RASTI...3..315S}
}

\end{document}